\documentclass{aa}  

\usepackage{graphicx}
\usepackage[varg]{txfonts}
\usepackage{siunitx}
\usepackage{natbib}
\usepackage{hyperref}
\hypersetup{colorlinks,citecolor=blue,linkcolor=blue}

\newcommand{\Nico}[1]{{#1}}
\newcommand{\name}[1]{\textit{#1}}
\newcommand{\software}[1]{\textit{#1}}
\newcommand{\minus}{\operatorname{-}}

\begin{document}

   \title{Variability of the slow solar wind: New insights from  modelling and \name{PSP-WISPR} observations.}
   \titlerunning{On the variability of the slow solar wind}
   \authorrunning{Poirier N. et al. (2023)}
   
    \author{Nicolas Poirier\inst{1,2}
    \and Victor Réville\inst{3}
        \and Alexis P. Rouillard\inst{3}
    \and Athanasios Kouloumvakos\inst{4}
    \and Emeline Valette\inst{3}}
    \institute{Rosseland Centre for Solar Physics, University of Oslo, Postboks 1029 Blindern, N-0315 Oslo, Norway \and
    Institute of Theoretical Astrophysics, University of Oslo, Postboks 1029 Blindern, N-0315 Oslo, Norway \and 
    IRAP, Université Toulouse III - Paul Sabatier, CNRS, CNES, Toulouse, France \and
    The Johns Hopkins University Applied Physics Laboratory, 11101 Johns Hopkins Road, Laurel, MD 20723, USA}

 
  \abstract
   {}
   {We analyse the signature and origin of transient structures embedded in the slow solar wind, and observed by the \name{Wide-Field Imager for Parker Solar Probe} (\name{WISPR}) during its first ten passages close to the Sun. \name{WISPR} provides a new in-depth vision on these structures, which have long been speculated to be a remnant of the pinch-off magnetic reconnection occurring at the tip of helmet streamers.}
   {We pursued the previous modelling works of \cite{Reville2020b,Reville2022} that simulate the dynamic release of quasi-periodic density structures into the slow wind through a tearing-induced magnetic reconnection at the tip of helmet streamers. Synthetic \name{WISPR} white-light (WL) images are produced using a newly developed advanced forward modelling algorithm that includes an adaptive grid refinement to resolve the smallest transient structures in the simulations. We analysed the aspect and properties of the simulated WL signatures in several case studies that are  typical of solar minimum and near-maximum configurations.}
   {Quasi-periodic density structures associated with small-scale magnetic flux ropes are formed by tearing-induced magnetic reconnection at the heliospheric current sheet and within $3-7\ \rm{R_\odot}$. Their appearance in WL images is greatly affected by the shape of the streamer belt and the presence of pseudo-streamers. The simulations show periodicities on  $\simeq90-180\ \rm{min}$, $\simeq7-10\ \rm{hr,}$ and $\simeq25-50\ \rm{hr}$ timescales, which are compatible with \name{WISPR} and past observations.}
   {This work shows strong evidence for a tearing-induced magnetic reconnection contributing to the long-observed high variability of the slow solar wind.}


   \maketitle
%

\section{Introduction}

In contrast to the fast solar wind, a mystery remains on the origin of the slow solar wind (SSW) and of its high variability. This variability can be the result of time-dependent and/or spatial-dependent effects. 

The spatial-dependent variability often emerges in structured bundles of bright rays in coronal white-light (WL) emissions, that have long been observed from coronagraphs and heliospheric imagers such as the \textit{Solar and Heliospheric Observatory} \citep[SoHO: ][]{Domingo1995} and the \textit{Solar TErrestrial RElations Observatory} \citep[STEREO: ][]{Kaiser2008}. Recently, the \name{Wide-Field Imager for Parker Solar Probe} \citep[WISPR: ][]{Vourlidas2016} revealed a finer structuring of the slow solar wind at scales down to the thin heliospheric plasma sheet (HPS) \citep{Poirier2020,Liewer2022,Howard2022}. 

On the other hand, it has been shown that the SSW is also highly time dependent, by hosting up to $\approx 80\%$ of quasi-periodic structures \citep{Viall2008}. This paper focus on the time-dependent variability of the slow wind, as captured from the \name{WISPR} novel perspective and in light of state-of-the-art modelling.\\ 

Density transient structures that propagate along with the SSW have long been observed in white-light imagery, with a great variety of shapes, speed, and origins. Among the most evident are coronal mass ejections (CMEs), which by releasing tremendous amount of coronal material into the heliosphere, generate significant brightness enhancements in both coronagraph and heliospheric images \citep[see e.g. the review by][]{Webb2012}. Some CME events that undergo more moderate and progressive accelerations,  called streamer blowouts, have been particularly observed to deflect towards the cusp of helmet streamers and further propagate within the SSW \citep[see e.g. the recent \name{WISPR} observations described in][]{Hess2020,Korreck2020,Rouillard2020b}.

Since the beginnings of the \name{SoHO-LASCO} coronagraph, other CME-like flux rope structures known as the Sheeley blobs \citep{Sheeley1997} have been observed to propagate along the bright rays associated with streamer stalks where the densest slow wind originates. Early interpretations suggested that these structures formed as a result of a pinch-off reconnection at the tip of helmet streamers that would have been stretched out beforehand \citep{Gosling1995,Wang1998}; the conditions leading to this stretching and eventually to the reconnection are still unclear. On some occasions they also appear as bright arches, which may be more or less `squashed' depending on their inclination with respect to the observer \citep[see e.g.][]{Sheeley2010}. 

Large loops from active regions (ARs) have also been observed to leave arch-like signatures as they gradually expand into the corona \citep[as seen in both X-ray and WL emissions; see ][]{Uchida1992,Morgan2013}. A helmet streamer made of such expanding loops may then be prone to stretching, and to the formation of streamer blobs via the pinch-off reconnection scenario.

This picture has the advantage of  also being consistent with observations of plasma inflows in \name{LASCO} \citep{Wang2000,Sheeley2002}, which were associated for the first time with outflowing blobs later on \citep{Sanchez-Diaz2017a,Lynch2020}. The continuous tracking of blobs expelled from the tip of helmet streamers all the way to points of in situ measurements reveals that they transport helical magnetic fields \citep{Rouillard2009,Rouillard2011a}, which is further supported by recent \name{Parker Solar Probe} (\name{PSP}) observations \citep{Lavraud2020,Rouillard2020a}. More systematic statistical analyses based on \name{STEREO} images, and of in situ measurements inside the HPS revealed that the topology of blobs is consistent with magnetic flux ropes \citep{Sanchez-Diaz2019} that could form via magnetic reconnection at the tip of helmet streamers \citep{Sanchez-Diaz2017b}. \\

The modelling of streamer instabilities in time-dependent magnetohydrodynamics (MHD) simulations also supports this scenario \citep{Chen2009,Lynch2020}. Following in the footsteps of these models, \citet{Reville2020b} investigated in detail the tearing instability that occurs near the cusp of streamers, in a high-resolution 2.5D simulation of the corona and using an idealistic dipolar configuration of the solar magnetic field. Streamer blobs were reproduced in addition to a plethora of quasi-periodic structures over a wide range of frequencies. Past studies based on near 1 AU remote-sensing \Nico{and/or} in situ observations also reveal the existence of quasi-periodic structures with periodicities varying from $\approx 90-180\ \rm{min}$ to $\approx 8-16\ \rm{hr}$ \citep{Viall2010,Viall2015,Kepko2016,Sanchez-Diaz2017a}, \Nico{which were found in Helios \citep{DiMatteo2019} and} in recent \name{PSP} observations as well \citep{Rouillard2020a}. From a high-cadence campaign on the \name{STEREO-A COR-2} coronagraph, \citet{DeForest2018} recently revealed the ubiquitous presence of density fluctuations at even smaller scales $\approx 20-40-60\ \rm{min}$. \\

In addition to the density and magnetic field, other plasma properties have also been measured to vary during the passage of these transients. For instance, \citet{Kepko2016} showed a similarity between the variability of the charge state ratios measured in situ in the slow wind and the short hourly timescale  of the quasi-periodic structures observed remotely in WL streamers. This SSW originating from streamers tends to exhibit high charge state ratios typical of hot ARs, whereas the SSW that emerges farther away from streamers, probably from deeper inside coronal holes is characterised by lower charge-state ratios comparable to those measured in the fast wind \citep{Neugebauer2002,Liewer2004,Stakhiv2015,Stakhiv2016}. The streamer-like SSW is also known to be more enriched in heavy ions having a low first ionisation potential (FIP) \citep[see e.g.][]{Steiger1996,Peter1998b}, a composition typical of closed-field plasma from ARs \citep[see e.g.][]{Ko2002,Brooks2011,Doschek2019}. The streamer-like SSW, or at least its dynamic component, could hence be conveniently interpreted as originating from the pinch-off reconnection mechanism  by offering a channel through which closed-field material can be intermittently released into the slow wind. \\

This paper further investigates this scenario through a qualitative comparison between the recent highly resolved WL observations taken by \name{WISPR}, and  high-resolution simulations of the solar corona and solar wind. We first analyse in section \ref{sec:WISPR_obs} two events observed by \name{WISPR} that depict quasi-periodic structures. We then present in Sect. \ref{sec:modeling} our modelling approach to reproduce such structures through the tearing-induced reconnection at the tip of streamers. Synthetic \name{WISPR} images are then produced and compared against observations in Sect. \ref{sec:results}. Limitations and future perspectives on this work are discussed in Sect. \ref{sec:discussion}. We finally conclude on the possible implications of this work for the understanding of the slow solar wind in Sect. \ref{sec:conclusion}.

\section{Observations}
\label{sec:WISPR_obs}

After its first 11 successful encounters \name{WISPR} has already provided a wealth of images rich in structures that were often unresolved from typical 1 AU observatories \citep[see e.g.][for an overview of the tenth encounter]{Howard2022}. This is a direct benefit of bringing an imager so close to the Sun, to a vantage point that is located inside the corona. By drastically shortening the line-of-sight integration path, \name{WISPR} is able to resolve with unprecedented detail the density structures that propagate within the slow solar wind. \name{WISPR} consists of two WL heliospheric imagers that are mounted on the ram side of \name{PSP}, and so the solar wind structures can be imaged prior to their in situ measurement \citep{Vourlidas2016}. \name{WISPR} offers a large field of view (FOV) thanks to its two telescopes that cover in elongation angle ($\epsilon$, angle away from the Sun) $13.5-53.0^\circ$ for the inner telescope (\name{WISPR-I}) and $50.5-108.5^\circ$ for the outer telescope (\name{WISPR-O}). At the closest approach to be reached by \name{PSP} in 2024 ($9.86\ \rm{R_\odot}$), \name{WISPR-I} will be able to observe the corona from only $2.3\ \rm{R_\odot}$. We exploit \name{WISPR} level 3 images,\footnote{data source: \url{https://wispr.nrl.navy.mil/wisprdata}} which have been calibrated \citep[see][]{Hess2021} and where contributions to the white-light emissions by dust particles (i.e. the F-corona) have been subtracted to reveal only the faint K-corona made up of coronal electrons \citep[see][for more details on the procedure]{Howard2022}. \\

\subsection{First insights on the transients observed by \name{WISPR}}

\begin{figure}[]
\centering
\includegraphics[width=0.48\textwidth]{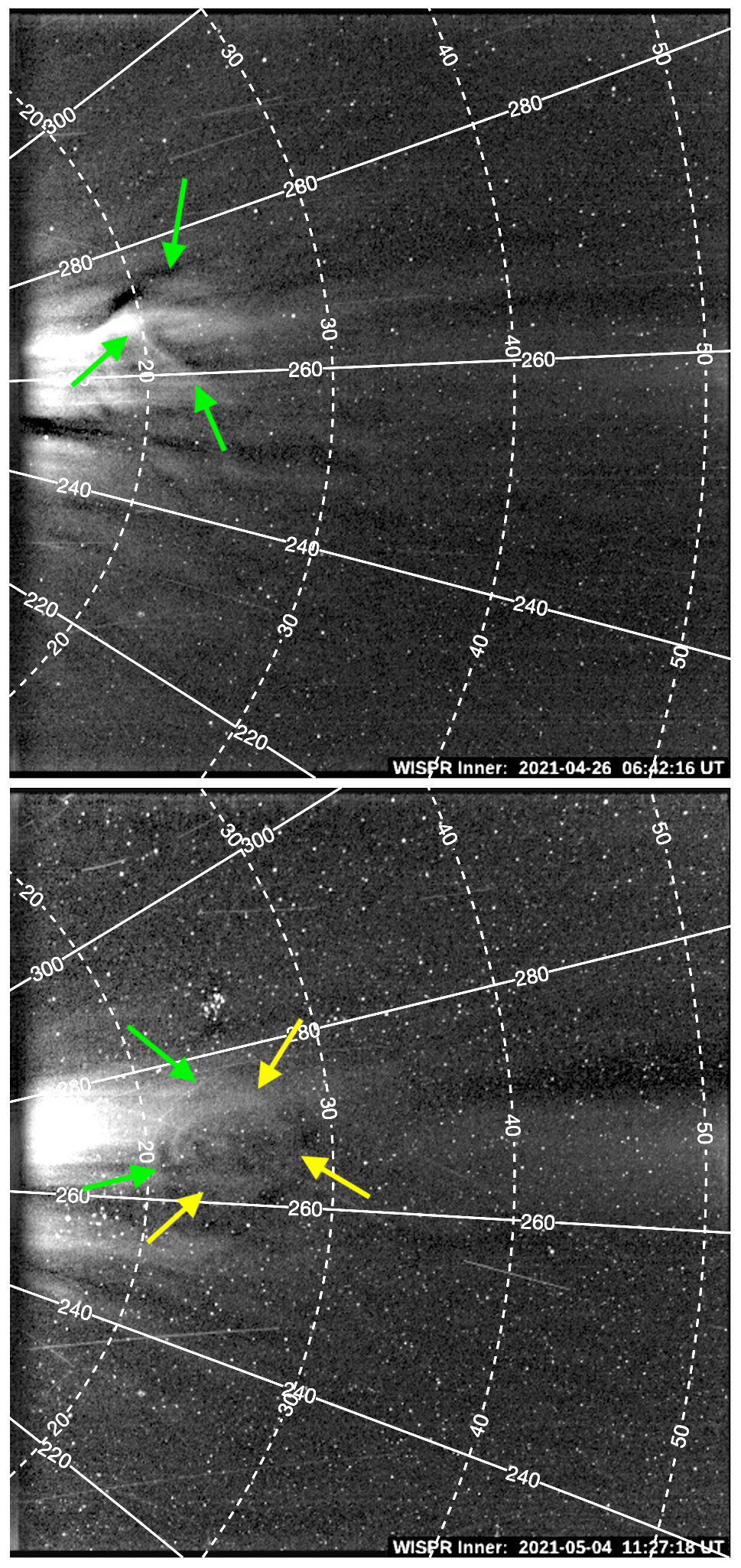}
\caption{Two examples of large-scale density structures observed by \textit{WISPR-I} on April 26, 2021, 06:42 universal time (UT, top panel) and May 04, 2021, 11:27 UT (bottom panel). The helioprojective-radial coordinate grid is plotted as solid white isolines for the position angle ($\psi$, i.e. the anticlockwise angle from solar north) and dashed white isolines for the elongation angle ($\epsilon$), both in degrees.
\label{fig:WISPR_blobs_large}}
\end{figure}

\textit{WISPR} has detected a wealth of fluctuations in the slow wind, whose signatures can be very diverse. Among these fluctuations many have been  associated with magnetic flux ropes with a clear dark cavity, suggesting that these flux ropes were likely   observed edge-on or at a small inclination angle. 

We present two examples of such signatures in Fig. \ref{fig:WISPR_blobs_large}, captured by the inner telescope \name{WISPR-I} during the eighth \name{PSP} encounter. Two bright shells (green arrows) can be seen, one   circular (bottom panel) and  the other   more V-shaped (top panel). Such shapes have already been observed from 1 AU, albeit to a larger spatial extent. They have been especially captured in great detail in events associated with pristine slow CMEs observed by \name{WISPR}. The V-shape was  interpreted either as a slight inclination of the flux rope with respect to the line of sight (LOS) of the observer or as a byproduct of the reconnection process itself that generates the flux rope \citep{Thernisien2006, Rouillard2009b,Rouillard2020b}. The fact that the brightness enhancement is often more marked at the back end of the flux rope supports the latter scenario, by an accumulation of plasma from the reconnection exhaust. A closer look at the bottom panel of Fig. \ref{fig:WISPR_blobs_large} shows an inner circular structure (yellow arrows) that has also been detected in other transients observed by \name{WISPR} \citep[see e.g.][]{Hess2020,Rouillard2020b,Howard2022}.

Most signatures of this scale have been found to be associated with sporadic (slow) CME events, where flux ropes are already present low in the corona well below the tip of streamers \citep[see e.g.][]{Hess2020,Korreck2020,Rouillard2020b,Howard2022}. In this paper we focus on flux ropes that form on a regular basis just above the tip of helmet streamers, which are presumed to be major contributors to the variability of the slow wind. \\

\begin{figure}[]
\centering
\includegraphics[width=0.48\textwidth]{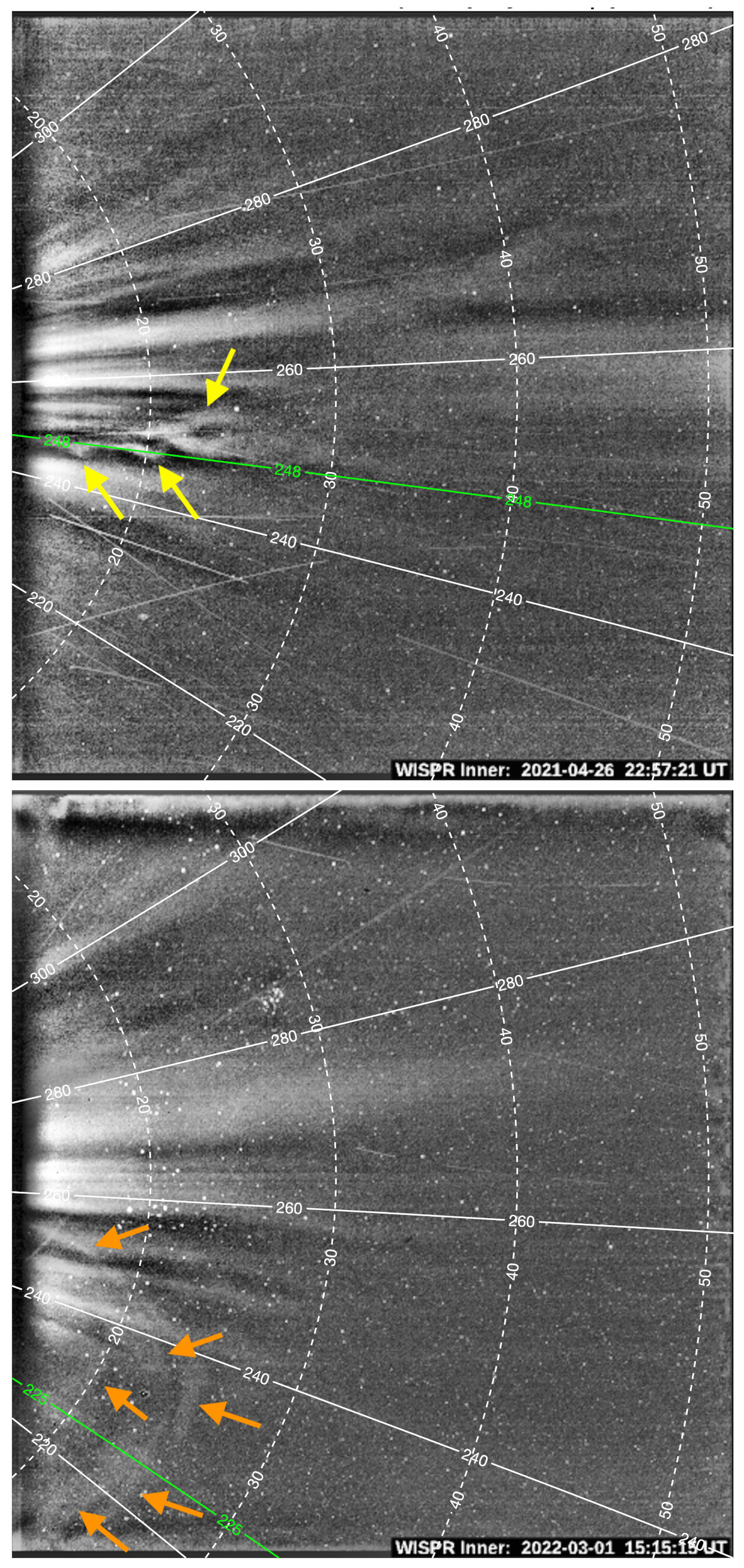}
\caption{Two examples of small-scale density structures observed by \textit{WISPR-I} on April 26, 2021, 22:57 UT (top) and March 01, 2022, 15:15 UT (bottom). The helioprojective-radial coordinate grid is again represented in the same format as in Fig. \ref{fig:WISPR_blobs_large}. The green lines represent the pixels extracted to build the time--distance (J-)maps shown in Fig. \ref{fig:WISPR_Jmaps}.
\label{fig:WISPR_blobs}}
\end{figure}

Compared to past near 1 AU observations, the novelty of \name{WISPR} observations is in imaging streamers from much closer in, providing clearer signatures of its embedded transients and access to smaller scales. In figure \ref{fig:WISPR_blobs} we show two events that may be related to quasi-periodic structures captured by \name{WISPR-I} during the 8th (top panel) and 11th (bottom panel) \name{PSP} encounters, hereafter referred to as the April 2021 and March 2022 events, respectively. We were able to identify a series of other similar events in \name{WISPR-I} images, but we only selected here the most visible ones for illustrative purposes, particularly in the most recent \name{WISPR-I} observations. Although more difficult to interpret because of a higher solar activity, they reveal local features that clearly stand out from the background signal.

The top panel unveils a track of three small-scale structures. Similarly to the larger-scale events shown previously in Fig. \ref{fig:WISPR_blobs_large}, they also appear as bright annuli suggesting flux ropes seen edge-on (\Nico{yellow} arrows). At this point it is hard to say whether these structures are actual small-scale flux ropes or if   they are located far away from \textit{WISPR}, which we discuss below. In contrast, the bottom panel shows arch-like signatures (orange arrows). Similar signatures have already been observed from 1 AU, and have been related to flux ropes seen with a greater inclination angle or almost face-on \citep{Sheeley2010, Rouillard2011a}, or also to expanding AR loops \citep{Morgan2013}. In the latter case though, the expansion of the loops is much slower than the propagation speed of the transients captured by \name{WISPR} (see the fitting performed in Sect. \ref{subsec:WISPR_obs_Jmaps}). 

Similarly to the April 2021 event shown in the top panel, a close-up visual inspection of the March 2022 event also reveals consecutive arches following each other. Both these events show interesting periodic behaviour in their spatial distribution, and hence they may be connected to the above-mentioned $90-180\ \rm{min}$ quasi-periodic structures that have been previously detected in the slow wind; this is  discussed further in Sect. \ref{subsec:WISPR_obs_Jmaps}. \\

Finally, we can give a rough estimate of the brightness variation induced by the passage of these transients. For this purpose, we examined the pixel values as given in units of mean solar brightness ($B_\odot$) in the level 3 \name{WISPR-I} \emph{.fits} files. It is important to note that these data products are not photometrically accurate because some of the K-corona emissions might be removed during the background removal procedure.\footnote{See the disclaimer about the level 3 (version 1) data at \url{https://wispr.nrl.navy.mil/wisprdata}} We averaged the emissions over representative areas that define the transients and the background (host) streamers, and computed the relative difference $(\Bar{B}_{transient}-\Bar{B}_{streamer})/\Bar{B}_{streamer}$. We found relative brightness increases of $\approx 80-95\%$ for the  April 2021 (edge-on case) event and $\approx 30-50\%$ for the March 2022 (face-on case) event, which  is brighter than what has been typically measured from 1 AU for the \citet{Sheeley1997} blobs. As we show  throughout this paper, the Sheeley blobs and the quasi-periodic structures captured by \name{WISPR} can be related to two different families of transients \Nico{that result from a} pinch-off reconnection at the tip of helmet streamers.

\subsection{Global context from near 1 AU observations}
\label{subsec:context_1au}

\begin{figure}[]
\centering
\includegraphics[width=0.5\textwidth]{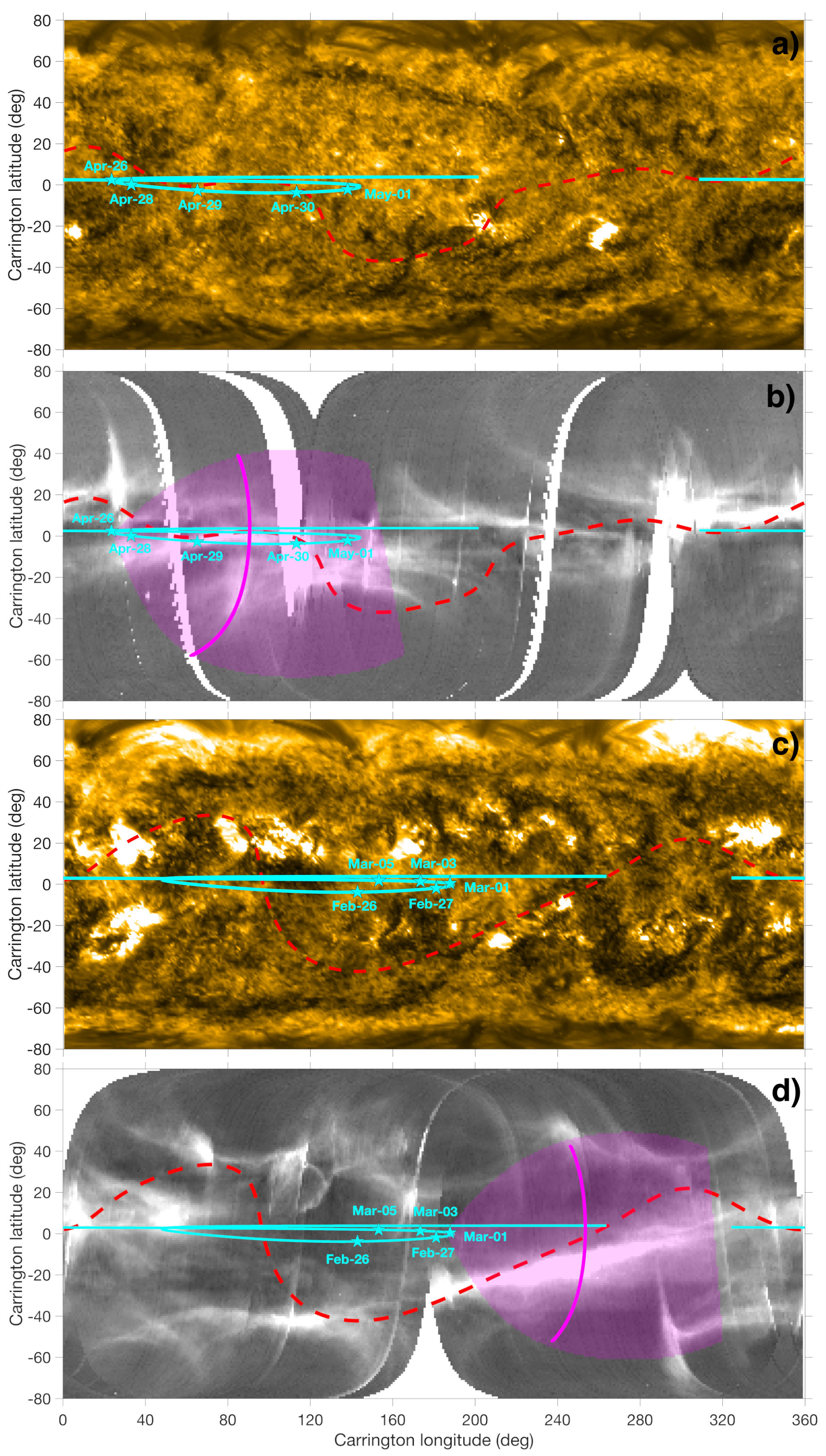}
\caption{ Contextual imagery from near 1 AU. Panels a and c: Synoptic maps of the low solar atmosphere derived from \name{SDO-AIA} (171\si{\angstrom}) EUV observations. Panels b and d: Synoptic maps of the WL corona as seen by the \name{SoHO LASCO-C2} coronagraph. These maps have been produced by combining on-disk (off-limb) images over a full (half) solar rotation prior to the April 26, 2021 (panels a and b), and March 01, 2022 (panels c and d), events \citep[see section 4.2 of][for a description of the method]{Rouillard2020}. The dashed red line represents an estimate of the shape of the HCS derived from PFSS extrapolations of the photospheric magnetic field. Among the multitude of photospheric maps available we picked two maps that match both the shape of the WL streamer belt \citep[see][]{Poirier2021} and the timings of the HCS crossings determined from magnetic sector measurements taken in situ at \name{PSP}. In the present case, the magnetic maps from the Air Force Data Assimilative Photospheric flux Transport (ADAPT) model \citep{Arge2010,Arge2011,Arge2013} for April 16 12:00UT (8th realisation) and March 01 12:00UT (11th realisation) were selected assuming a source surface height of $2.0\ \rm{R_\odot}$ and $2.5\ \rm{R_\odot}$, respectively. The vertical (radial) projection of \name{PSP} orbit onto the solar disk is plotted in cyan; the  stars  indicate several dates at \name{PSP}. The magenta surfaces represent the projections of the regions scanned by \name{WISPR-I} at the time of the two images shown in Fig. \ref{fig:WISPR_blobs}, with a solid line indicating the location of the Thomson sphere extracted at $\epsilon=22^\circ$. \label{fig:C2_maps}}
\end{figure}

To get a better context for these events, we construct WL maps of the streamer belt as observed from near 1 AU by \name{LASCO-C2} over half a solar rotation. We show these maps in Fig. \ref{fig:C2_maps} (panels b and d). Estimates of the heliospheric current sheet (HCS) derived from potential field source surface (PFSS) extrapolations are plotted (red dashed lines) to help us differentiate pseudo-streamers (unipolar structures) from the main streamer belt (where the magnetic polarity switches sign). Assuming that the above transients originate from and propagate within the streamer belt, we  identified two possible source regions that have an inclination consistent with the April 2021 and March 2022 events observed by \name{WISPR}. Since the imaged transients significantly stand out from the background streamers, they should be located quite close to the Thomson sphere (see the magenta lines) where most of the WL emissions are expected to originate (see Sect. \ref{subsec:synthetic}). We identified two possible source regions, a nearly aligned section of the streamer belt located within $60-110^\circ$ (and $\approx 0^\circ$, see panel b) of Carrington longitudes (and latitudes), and an inclined section of the streamer belt located within $220-280^\circ$ (and $\minus20-0^\circ$, see panel d). \Nico{In an independent study \citet{Liewer2023} determined a similar source region for the April 2021 event using a more sophisticated tracking technique, and hence supports the fact that this event was indeed propagating within the main streamer belt.} \\

The low coronal structures underlying the streamer belt, as observed in the  extreme ultraviolet (EUV) by the Atmospheric Imaging Assembly (\name{AIA})   on board the Solar Dynamics Observatory (\name{SDO}) are shown in panels a and c of Fig. \ref{fig:C2_maps}. In panel c two major ARs can be clearly seen at longitudes $210^\circ$ and $245^\circ$, which are located underneath the potential source streamer that produced the March 2022 event. These ARs may prove important in the formation of streamer transients. The hot plasma in such active closed-field regions can be prone to expansion into the corona through thermal instability, bringing the AR loops that are frozen into that plasma to higher coronal heights, eventually up to the cusp of streamers. On the other hand, in panel a no significant AR is visible beneath the streamer that potentially produced the April 2021 event observed by \name{WISPR}. However, it has also been suggested that the stretching of streamers can naturally occur simply as the result of the magnetic field near the cusp being too weak to hold the thermal pressure exerted by the underlying plasma \citet{Chen2009}.

The two mechanisms could act   together in the formation of streamer transients through the pinch-off reconnection process, which is actually supported by   observations \citep[see e.g.][]{Morgan2013} and by simulations \citep{Chen2009,Reville2020b}. In particular, as we discuss in the results section (\ref{subsec:results_2D}), that the pinch-off reconnection process can generate transients with a low frequency that is quite variable and dependent on the local coronal conditions beneath the streamer. Therefore, the presence and amount of ARs beneath streamers could implicitly affect the rate at which these low-frequency transients are produced. From past observations near 1 AU, these transients were detected with periods varying from $\approx 8-16\ \rm{hr}$ \citep[][near solar maximum]{Sanchez-Diaz2017b} to $0.5-2\ \rm{days}$ \citep[][near solar minimum]{Morgan2021}. A future statistical study that links these heliospheric measurements to low atmospheric EUV observations would be helpful to better assess the contribution of ARs on the release of streamer transients.

\subsection{Tracking transients in \name{WISPR} J-maps}
\label{subsec:WISPR_obs_Jmaps}

The usual and efficient method to further characterise transient features and have more insights into their possible generation mechanism is to measure their periodicity. For this purpose, a method that has been widely used across the community is to track transients in distance-time maps called J-maps \citep[see e.g.][]{Sheeley1999,Sheeley2008,Rouillard2008,Rouillard2009}. The bright features that appear in  J-maps then provide insightful information on the periodicity, propagation speed, and acceleration profiles of the  transients (albeit with some limitations, as discussed below). \\

These maps are commonly produced by extracting pixels along a fixed direction (either at the ecliptic or at another position angle), and using the elongation angle ($\epsilon$) as a measure of the angular distance away from Sun centre. 

Tracking transient features in heliospheric images has long been a delicate task, and even more for a rapidly moving and up-close imager like \name{WISPR}. New techniques have been developed to better track \name{WISPR} features, whether they are static \citep[e.g. coronal rays:][]{Liewer2022} or dynamic \citep[e.g. CMEs:][]{Liewer2020}. These techniques include a number of corrections to account  for the effects of spacecraft motion, perspective, and orbit out of the solar equatorial plane. For instance, a transient propagating radially outwards from the Sun does not necessarily remain at a constant position angle as it moves across the \name{WISPR} FOV, and its distance to \name{WISPR} may also vary. This can affect their appearance in J-maps. For instance, curved signatures were noted during a CME event that came close to the two heliospheric imagers on board \name{STEREO} \citep{Sheeley2008, Rouillard2008}. Furthermore, when a target moves away from the observer, hence leading to an apparent slowing of its propagation speed, this can also produce curved signatures in J-maps as we see later for the April 2021 event. Performing a precise fitting of the transients observed by \name{WISPR} is beyond the scope of this study, and hence circumvents the need of a complex tracking technique as developed by \citet{Liewer2022}.\\

\begin{figure*}[]
\centering
\includegraphics[width=0.99\textwidth]{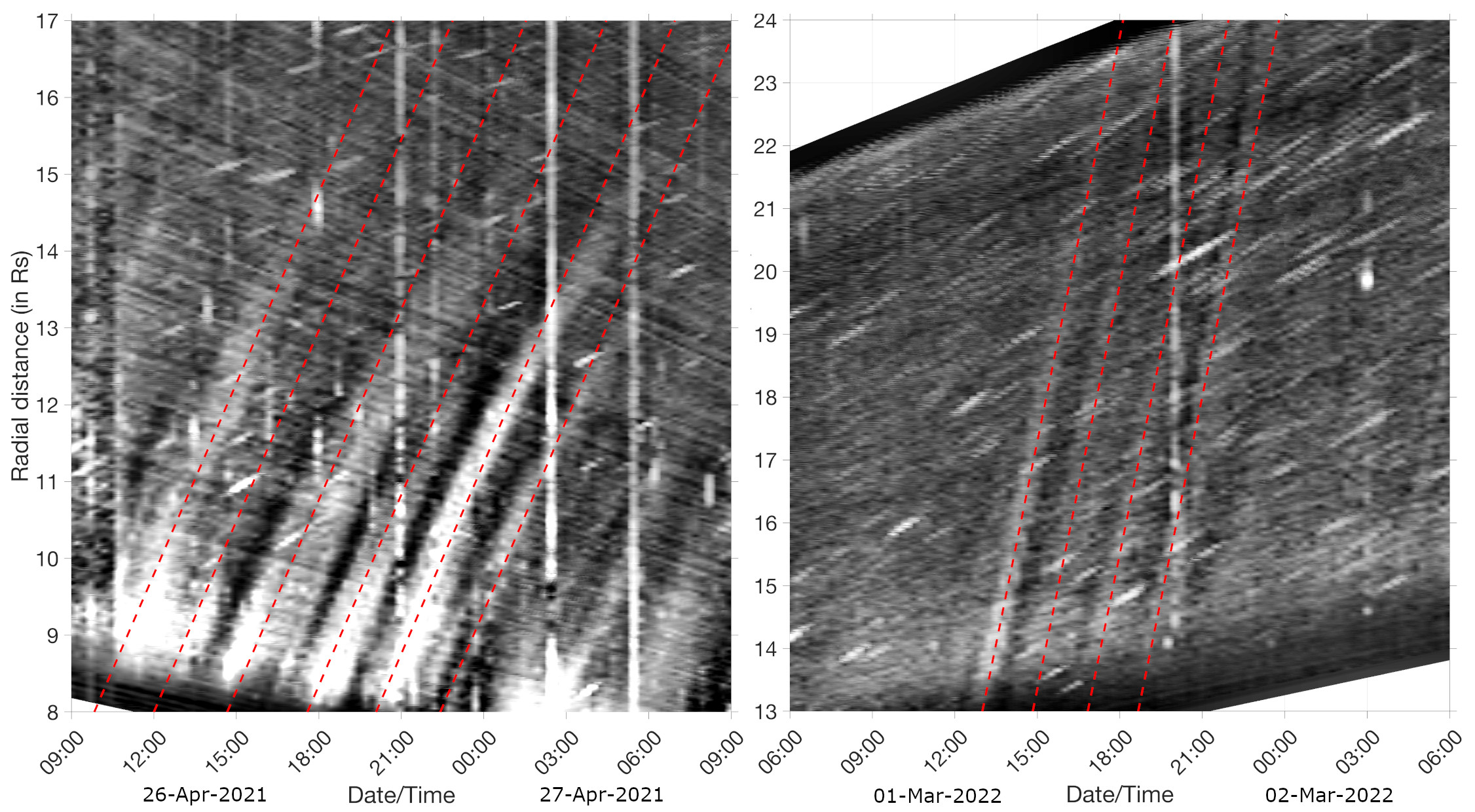}
\caption{ \name{WISPR-I} distance--time (J-)maps for the April 2021 (left) and March 2022 (right) events, and along the two slits shown in Fig. \ref{fig:WISPR_blobs} (solid green lines). The fitted profiles are plotted as dashed red lines, with constant velocity of $160\ \rm{km/s}$ (left) and $415\ \rm{km/s}$ (right), i.e. about the solar wind bulk velocity measured a few days later by \name{PSP-SWEAP/SPC} in situ (see Fig. \ref{fig:PSP_insitu}).
\label{fig:WISPR_Jmaps}}
\end{figure*}

The J-maps associated with the March 2022 and April 2021 events introduced earlier are shown in Fig. \ref{fig:WISPR_Jmaps}. The corresponding slits along which the J-maps were constructed are plotted in Fig. \ref{fig:WISPR_blobs} as green lines. These J-maps are classical distance--time maps, except that instead of the typical elongation angle that measures the angle away from the  Sun centre, the radial distance to the Sun is used. The implicit assumption to determine this parameter, however, is that extracted pixels are projected onto the Thomson sphere, a reference surface where WL emissions are expected to be strongest (see discussion in Sect. \ref{subsec:synthetic}). From these J-maps we were able to infer some insightful properties of the transient structures captured by \name{WISPR}, which we describe in the following paragraphs.  \\

\noindent \textit{\textbf{Propagation profiles}}\\
The J-maps allow us to perform a direct visual inspection of the speed profile of the propagating structures. For this purpose we fitted profiles of constant speed for several of the most visible stripes \Nico{in Fig. \ref{fig:WISPR_Jmaps}}. We note that several sub-structures could also be seen   between some of the brightest fronts, but were too faint to be shown here. The March 2022 event (right panel) is fairly well described by constant speed profiles at $\approx 415\ \rm{km/s}$. In contrast, the April 2021 event (left panel) shows curved stripes that deviate from constant speed profiles, and with a very low speed ($\approx 160\ \rm{km/s}$). \Nico{Using a more sophisticated tracking technique on the April 2021 event, \citet{Liewer2023} obtained a similar low propagation speed of $\approx 190\ \rm{km/s}$.} Although such a low propagation speed could be due to perspective effects (e.g. of the transients moving away from \name{WISPR}), we show in Sect. \ref{subsec:WISPR_insitu} that here this is probably  associated with a very slow and dense slow solar wind flow within the HPS.

Curved signatures in J-maps have already been found in \name{STEREO} observations for instance, \citep{Sheeley2008,Rouillard2008} and also more recently in \name{WISPR} \citep{Howard2022}. These apparent decelerations were in fact associated with the effect of the imaged structures getting closer to or farther away from the observer. Since \name{PSP} remained relatively `static' (i.e. co-rotating with the solar corona) at that time, that might suggest that the structure itself was moving with respect to \name{WISPR}. This is also consistent with the WL signatures that suggest \Nico{that these} flux ropes \Nico{propagated} along \name{PSP} orbital plane (see the green dashed line in the top panel of Fig. \ref{fig:WISPR_blobs}), where \Nico{their legs} may have come closer to or farther away from \name{WISPR} during their expansion. The pinch-off reconnection mechanism at the tip of streamers indeed predicts that such flux ropes develop large azimuthal extents, from their generation and during their expansion in the solar wind \citep{Sanchez-Diaz2019}. \\

\noindent \textit{\textbf{Periodicities}}\\
The periodicities measured between the fitted stripes range from $\approx 110-120\ \rm{min}$ (\Nico{Fig. \ref{fig:WISPR_Jmaps},} left panel) and $\approx 130-175\ \rm{min}$ (right panel)  between the fitted stripes. This falls well within the typical $\approx 90-180\ \rm{min}$ range previously detected from near 1 AU observations. Furthermore, similar periodicities are also retrieved in an upcoming statistical study from \citet{Viall2023} that includes a few \name{PSP} encounters. Because of the rapidly varying viewing conditions of \name{WISPR}, quasi-corotation phases do not last very long, and hence periodicities above $\approx 10\ \rm{hr}$ cannot easily be detected. Therefore, it remains difficult to check whether the longer $\approx 8-16\ \rm{hr}$ periods of streamer blobs measured previously from slowly moving 1 AU observatories also manifests in \name{WISPR} images.

As we  discuss in the next section, these periodicities could be byproducts of the pinch-off reconnection process occurring at the tip of streamers. More precisely, they could be  the manifestation of multiple modes associated with the tearing instability that can develop at the HCS.

\subsection{Insights from plasma measurements taken in situ at \name{PSP}}
\label{subsec:WISPR_insitu}

\begin{figure*}[]
\centering
\includegraphics[width=0.95\textwidth]{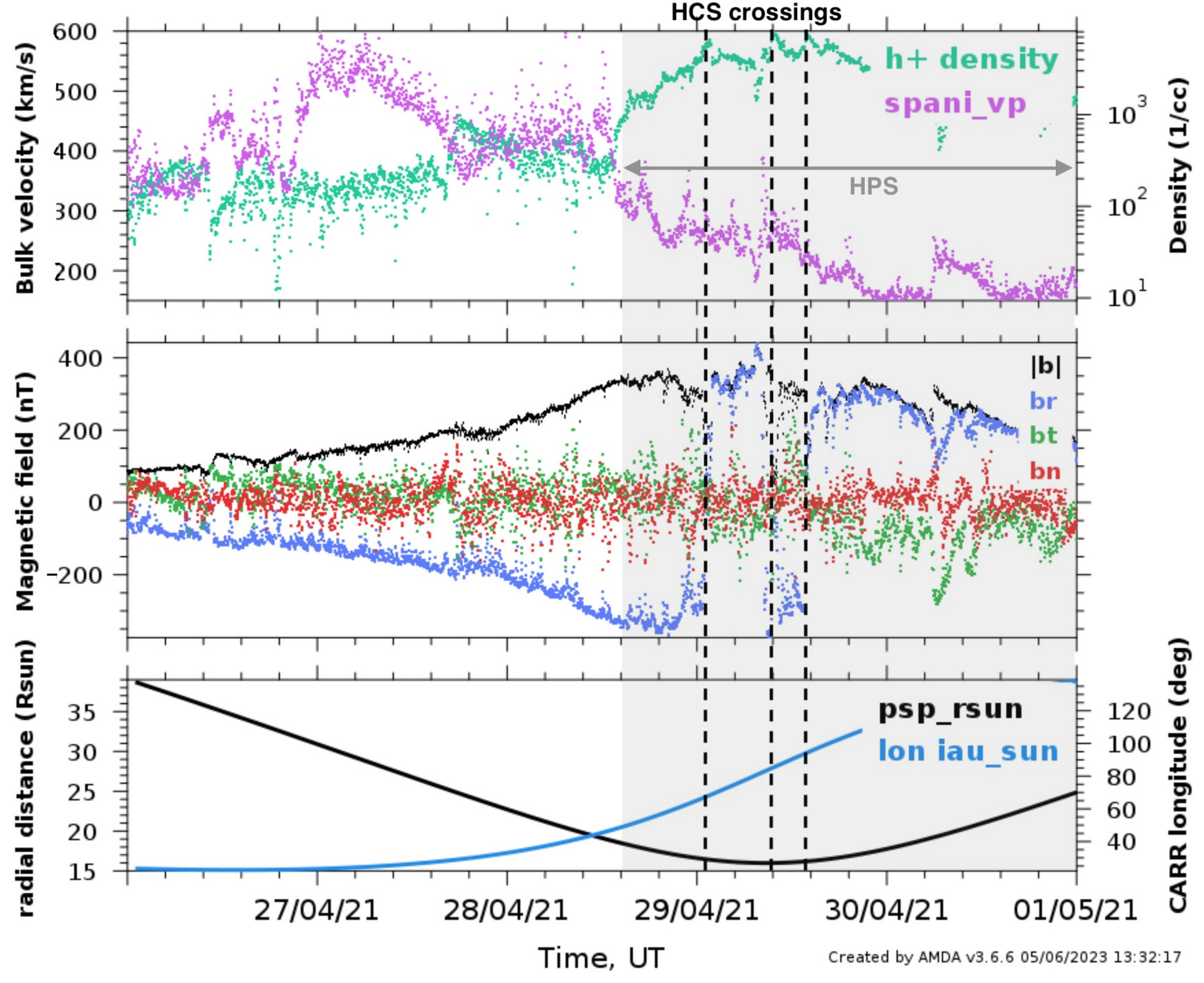}
\caption{ Solar wind measurements taken in situ at \name{PSP}. Top panel: Bulk velocity (left axis) and density of the proton solar wind \Nico{(right axis)} measured by the \name{Solar Probe Analyzer} (\name{SPAN-i}), which is part of the \name{Solar Wind Electrons Alphas and Protons} \citep[SWEAP:][]{Kasper2016} investigation. Middle panel: Magnetic field (1 min average) measured by the magnetometer (\name{MAG}) of the \name{FIELDS} instrument suite \citep{Bale2016}. Bottom panel: Radial distance to the Sun and Carrington longitude of \name{PSP}. A few days after the April 2021 event observed by \name{WISPR-I}, \name{PSP} sampled a very dense and slow solar wind typical of HPS, with several magnetic polarity inversions suggesting potential crossings of the HCS. Figure produced with the \name{AMDA} web tool publicly available at (\url{http://amda.irap.omp.eu/}).
\label{fig:PSP_insitu}}
\end{figure*}

We check whether the low propagation speeds measured for the April 26, 2021, event imaged by \name{WISPR-I} are realistic, by making a rough comparison with the solar wind speeds measured in situ at \name{PSP} around that time, as shown in Fig. \ref{fig:PSP_insitu}. This is possible because after that event, \name{PSP} entered in a pro-grade phase, and hence could a few days later sample a solar wind channel that probably hosted the transients captured by \name{WISPR-I} (see the \name{PSP} orbit plotted in the top panel of Fig. \ref{fig:C2_maps}). 

In the top panel, we can see that \name{PSP} indeed measured very slow (purple dots) and dense (emerald green dots) solar wind with $\approx 160-250\ \rm{km/s}$ and $\approx 1-8\times 10^3\ \rm{1/cc}$ at around $15\ \rm{R_\odot}$. Such plasma flows are typical of coronal streamers at that distance \citep{Cho2018,Morgan2020}, and more generally of HPS \citep[see e.g.][]{Winterhalter1994,Sanchez-Diaz2019, Lavraud2020}, which is also supported by a few HCS crossings during this interval (here simply identified by a global polarity inversion of the magnetic field, see middle panel). These SSWs are hence potential hosts of blobs and quasi-periodic structures like those observed by \name{WISPR-I}. Unfortunately, in this case the transients imaged by \name{WISPR-I} on April 26 (located within $\approx 8-16\ \rm{R_\odot}$) already moved far away before \name{PSP} could get inside that streamer belt starting from April 29. Nevertheless, the solar wind velocities measured there (at $\approx 15\ \rm{R_\odot}$, see the top and bottom panels of Fig. \ref{fig:PSP_insitu}) closely match   the $\approx 160\ \rm{km/s}$ fitted speed of the imaged transients. \\

\section{Modelling: Method}
\label{sec:modeling}

Several main characteristics of transients in the slow wind have been extracted from \name{WISPR} observations. We introduce in this section our modelling approach to get more insights into the possible origin of these structures. We  test the idea that the pinch-off reconnection process at the tip of streamers is responsible for the formation and release of the small transients observed by \name{WISPR}. 

This mechanism is tested first in an idealistic simulation of a very high-resolution time-dependent 2.5D  MHD dipolar corona (Sect. \ref{subsec:model_2D}), and then a lower resolution time-dependent 3D MHD simulation of the conditions encountered by \name{PSP} during its ninth passage near the Sun (Sect. \ref{subsec:model_3D}). We then present in Sect. \ref{subsec:synthetic} our approach to building synthetic products that can be compared with \name{WISPR} observations.

\subsection{Idealistic simulation of a dipolar corona}
\label{subsec:model_2D}

\begin{figure*}[]
\centering
\includegraphics[width=0.8\textwidth]{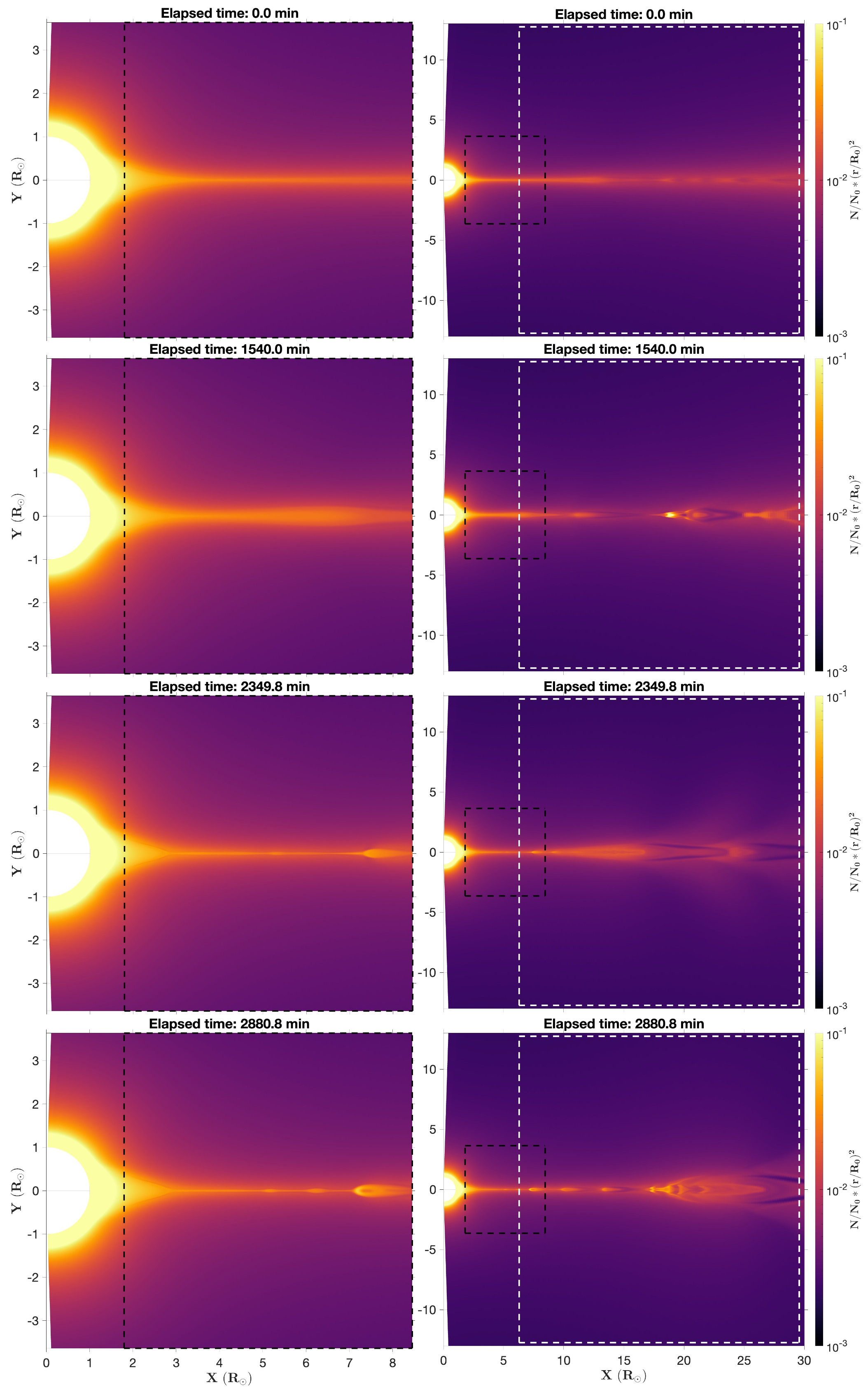}
\caption[]{ Simulated electron density from the idealistic dipolar set-up. Colours are plotted in a logarithmic scale. The approximate FOV of \name{WISPR-I} when \name{PSP} is located at $10\ \rm{R_\odot}$ and $35\ \rm{R_\odot}$ is depicted by a dark and white dashed rectangle, respectively. An animated version of this figure is available online at \url{https://doi.org/10.5281/zenodo.8135596}.
\label{fig:PLUTO2D}}
\end{figure*}

\citet{Reville2020b} describe in detail the pinch-off reconnection mechanism induced by the tearing instability, in an idealistic 2.5D simulation of the solar corona. To allow a fair comparison with actual observations from \name{WISPR}, the \citet{Reville2020b} simulation was run again with more outputs (one every $\simeq13\ \rm{min}$) to match the typical temporal cadence of \name{WISPR}.

In figure \ref{fig:PLUTO2D} we illustrate the main phases of the pinch-off reconnection mechanism, with several snapshots extracted from the simulation and zoomed-in views in the panels on  the left side. Starting from a near equilibrium state ($t=0\ \rm{min}$, first row), the tip of the helmet streamer eventually expands ($t=1540\ \rm{min}$, second row) due to pressure imbalance between the closed-field plasma confined beneath (as in coronal loops) and the out-flowing plasma from the adjacent open field. As the helmet streamer expands a thinning also occurs at its back end, up to the point where the streamer gets sufficiently thin locally for the tearing instability to trigger magnetic reconnection ($t=2349.8\ \rm{min}$, third row), referred to as the `ballooning mode' in \citet{Reville2020b}. The ejecta of a large plasmoid of dense and initially closed-field material follows ($t=2880.8\ \rm{min}$, fourth row). The tearing instability further develops at smaller scales triggering reconnection at multiple secondary sites, and is called the `tearing mode'. Plasma material is pushed away from these reconnection sites and then accumulates in small-scale and dense plasmoids. More precisely, this plasma concentrates in shell-like structures where the magnetic field is mostly poloidal, as in magnetic flux ropes. In contrast, the core of these structures is less dense due to a dominant toroidal magnetic field component that is  directed across the figure plane (see  animation associated with Fig. \ref{fig:PLUTO2D} available online,\footnote{\url{https://doi.org/10.5281/zenodo.8135596}} as well as those presented in the original paper by \citealt{Reville2020b}).

For the first time, the inner and outer \name{WISPR} telescopes combined can provide an in-depth view of the transient structures that form from pinch-off reconnection, right in their formation regions. As illustrated in figure \ref{fig:PLUTO2D}, \name{WISPR} may see different signatures according to its distance from the Sun, where the dashed white and dark rectangles show the approximate \name{WISPR-I} FOV, assuming that \name{PSP} is located respectively at $35\ \rm{R_\odot}$ and $10\ \rm{R_\odot}$ from the Sun  ($10\ \rm{R_\odot}$ being an estimate of the closest approach to be reached by 2024). For instance, it happens that some of the simulated transients eventually merge together along their propagation to form larger and/or denser plasmoids (see lower left panel of Fig. \ref{fig:PLUTO2D}). Hence, we pursue here the work of \citet{Reville2020b} to examine how such simulated structures may look in a white-light imager such as \name{WISPR}. \\

To produce synthetic white-light observables, we need first to extend the 2.5D simulated domain into three dimensions. We  perform an axisymmetric demultiplication of the 2.5D simulation about the solar rotation axis, hence producing a 3D corona with a flat streamer belt at the equator. By changing the position of our virtual observer we can then test most situations encountered by \name{WISPR} along its orbits and more generally throughout the solar cycle, that is from a horizontal to vertical streamer belt configuration typical of a solar minimum and maximum, respectively.

\subsection{Case study simulation of the ninth \name{PSP} encounter}
\label{subsec:model_3D}

Because the 2.5D set-up presented in \citet{Reville2020b} is highly idealistic, an attempt has been made to extend this work to a fully fledged 3D model that is called WindPredict-AW \citep[see][for a detailed description]{Parenti2022,Reville2022}. In this modelling framework, the 2.5D magnetic structures mentioned above translate into 3D magnetic flux ropes, where their generation and propagation can now be studied in a self-consistent manner. The disadvantage, however, is that the 3D set-up \Nico{provides a lower level of spatial resolution than the 2.5D set-up}. Since magnetic reconnection is allowed by numerical diffusion of the numerical scheme itself, it is bound by the actual numerical size of the mesh near the HCS. The 3D set-up is hence not optimal for the full development of the tearing instability, as is the idealistic set-up \citep[see][for more details]{Reville2022}. Despite this limitation, the 3D simulation still does produce transients but at low frequency, which are the ballooning modes and only the quasi-periodic structures with periods $\gtrsim 4\ \rm{hr}$. By applying a realistic photospheric magnetic map at the inner boundary, \citet{Reville2022} was able to reproduce the statistical occurrence of streamer flux ropes that intersected both \name{PSP} and \name{SolO} during the joint observation campaign of June 2020. \\

We here pursue the work of \citet{Reville2022} with a similar 3D simulation set-up, but applied to the ninth \name{PSP} encounter (August 2021). The inner boundary is set with the GONG-ADAPT (11th realisation) magnetogram of August 14, 2021, 00:00 UT, and kept fixed over the entire simulated period. The magnetogram was selected among many different sources and dates to best match the observed shape and location of the streamer belt, as seen from 1 AU by \name{SoHO-LASCO} over a full solar rotation. The selection process is based on the method presented in \citet{Poirier2021}. Another criteria was also the correct prediction of both magnetic sectors and timing of HCS crossings measured in situ by \textit{PSP}. Once the simulation relaxed, outputs of the entire 3D simulated domain were extracted every $\simeq 13\ \rm{min}$ to match the actual cadence of \textit{WISPR}. For the sake of computational time the simulation was run until $100$ outputs were obtained, which covers a time interval of $\simeq 22\ \rm{hr}$ starting at perihelion. The simulation is kept fixed outside this interval, and allows us to synthesise \name{WISPR} images over a longer period even though the simulated solar wind remains static. The static phase is still meaningful to differentiate the effect of the fast-moving probe from the propagation of the solar wind structures within the synthesised images. The procedure to produce \name{WISPR} synthetic images is described in Sect. \ref{subsec:synthetic}. \\

\begin{figure*}[p]
\centering
\includegraphics[width=0.75\textwidth]{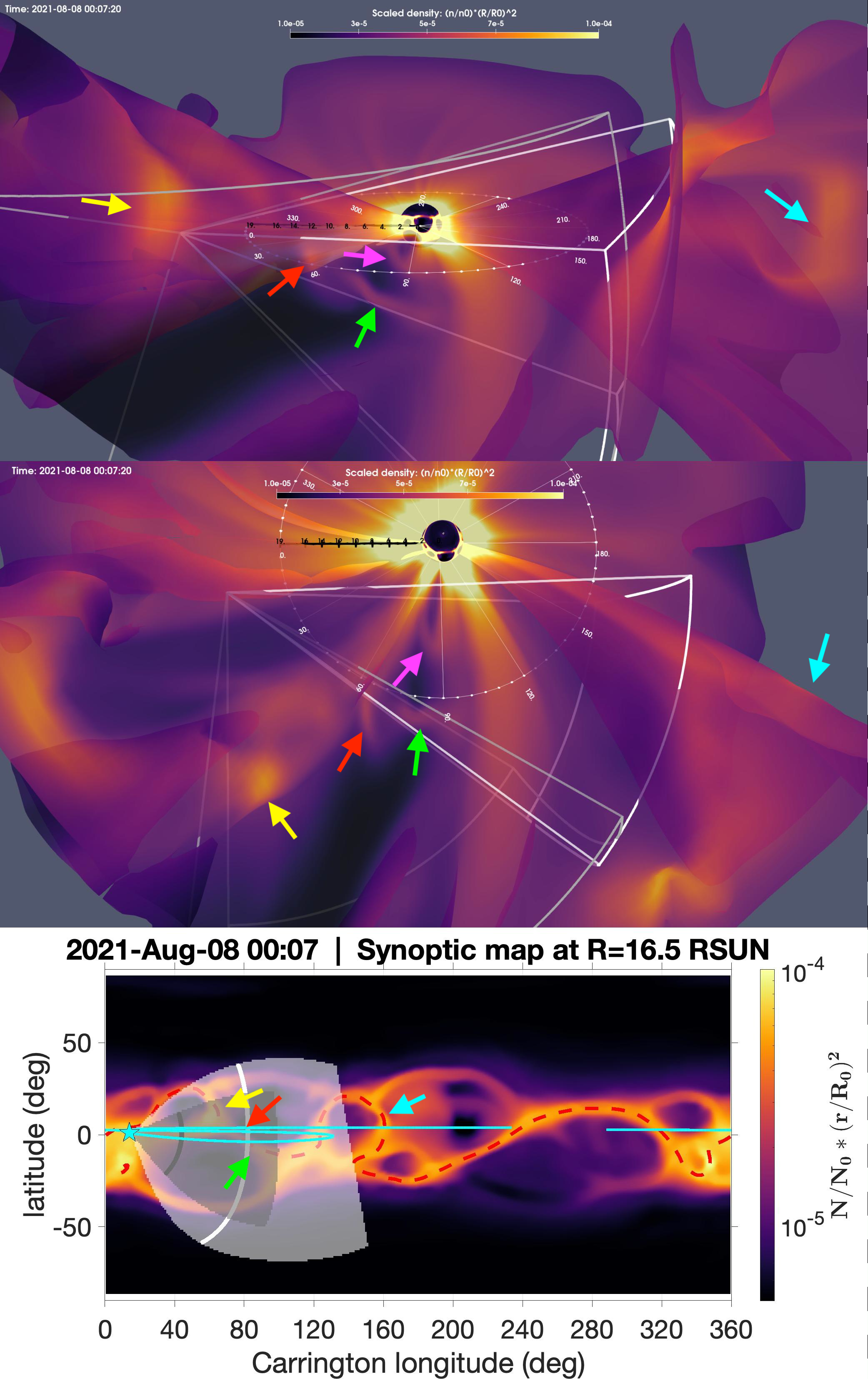}
\caption[]{ Simulated density structures from the WindPredict-AW fully fledged 3D MHD model. Top two  panels: Front and top views of the 3D density structures propagating within the core of the (true) streamer belt (identified with a $Br=0$ isosurface). Coordinates in the Carrington frame are also given, for the longitudes (lighter colours, in degrees) and radial distances (darker colours, in solar radii). Bottom panel: Synoptic (Carrington) map at $r=16.5\ \rm{R_\odot}$ with \name{PSP} orbit plotted in cyan. To identify the true streamer belt from pseudo-streamer structures, a $Br=0$ contour is plotted as a red dashed line. In both panels the FOV of \name{WISPR-I} and \name{WISPR-O} are shown in white and grey,  respectively. In both panels the logarithmic colour scale represents the flux of plasma density scaled by its value at the inner boundary ($r=1.0002\ \rm{R_\odot}$). The coloured arrows point to flux rope structures (see main text for dicussion). An animated version of this figure is available online at \url{https://doi.org/10.5281/zenodo.8135596}.
\label{fig:PLUTO3D}}
\end{figure*}

The simulated streamer belt and density structures propagating within its core are shown in Fig. \ref{fig:PLUTO3D}, along with the FOV of both \name{WISPR-I} (in white) and \name{WISPR-O} (in grey). At that time \name{WISPR} was imaging from a distance of $\approx 26\ \rm{R_\odot}$, a highly warped streamer belt typical of a high solar activity. Throughout the region scanned by \name{WISPR}, the streamer belt undergoes significant latitudinal shifts within $\approx \minus15-25^\circ$ of Carrington latitude. A few flux rope structures have been identified in the simulation (see the coloured arrows). All of them, except the  farthest one (cyan arrow)   produce visible WL signatures in the synthetic \name{WISPR} images (see   Sect. \ref{subsec:results_3D}). The flux ropes have different extents and widths that can be explained by a different stage of their formation and/or evolution. We also find  that the spatial extent of these flux ropes within the streamer belt varies, and that it is delimited by intersections of pseudo-streamers with the main streamer belt \citep[see also][]{Reville2022}. This makes up a complex network that is inherently connected to the   S-web \citep[or web of separatrix and quasi-separatrix layers; see][]{Antiochos2011}. Finally, each of these flux ropes shows a different inclination. All of this will affect their appearance from the \name{WISPR} perspective as we show in Sect. \ref{subsec:results_3D}.

\subsection{Producing synthetic \name{WISPR} images}
\label{subsec:synthetic}

Synthetic \name{WISPR} images are produced similarly to what was done in \citet{Poirier2020}, except that in the present work we used time-dependent simulations rather than   static simulations. \\

Following the Thomson scattering theory \citep{Howard2009,Howard2012}, the total intensity received by a pixel detector from scattered electrons can be expressed as an integral along the path length $z$ along each LOS:
\begin{equation}
\begin{aligned}
    I^{tot}_t &= \int_{z=0}^{z\rightarrow +\infty} I_t dz=\int_{z=0}^{z\rightarrow +\infty} n_e z^2 G dz\ \quad \text{(in }\rm{W.m^{-2}.sr^{-1}}\text{)} \\
    G &= \frac{B_\odot\pi\sigma_e}{2z^2}\left(\underbrace{2}_{1}\underbrace{\left[(1-u)C+uD\right]}_{2}\underbrace{-\sin{\chi}^2}_{1}\underbrace{\left[(1-u)A+uB\right]}_{2}\right)
\end{aligned}.
\label{eq:Thomson_scattering}
\end{equation}
Here $I_t$ refers to the total (and not polarised) intensity, $B_\odot\simeq 2.3\times 10^7\ \mathrm{W.m^{-2}.sr^{-1}}$ the Sun's mean radiance (or surface brightness), and $\sigma_e=r_e^2\simeq 7.95\times 10^{-30}\ \mathrm{m^2}$ the electron cross-section. The electron density $n_e$ is an input 3D datacube interpolated at each LOS point. The $G$ function here includes contributions from both pure-geometric scattering (indicated by $1$) and the solar illumination function (indicated by $2$). For far distances to the Sun, $G$ can be approximated as $(R_\odot/r)^2(2-\sin{\chi}^2)=(R_\odot/r)^2(1+\cos{\chi}^2)$, where $\chi$ is called the scattering angle between the scattering site and the Sun-observer line, and $(R_\odot/r)^2$ represents the classical fall-off of sunlight with heliocentric distance $r$ \citep[see][]{Howard2012}. For an observer as close to the Sun as \name{WISPR},  additional effects should be considered,  such as the collimation of sunlight and limb-darkening, using for instance the \name{van de Hulst} coefficients $A$, $B$, $C$, and $D$ defined in \citet[][Eqs. 25-28]{Howard2009}. A direct observation of Eq. (\ref{eq:Thomson_scattering}) shows that the integral is semi-infinite on the path length $z$. In practice, it is possible to  shrink this integral to a limited (finite) region that includes most contributions to the total brightness (see  discussion in Appendix \ref{sec:numerical}).

Theoretical works have shown that   WL emissions produced from Thomson scattering are expected  to peak at a surface called the Thomson sphere \citep[TS,][]{Vourlidas2006}. This can be geometrically defined by a sphere with its centre located halfway along the Sun--observer line, and with the length of this line for diameter. However, \citet{Howard2009} and \citet{Howard2012} have demonstrated that this peak at the Thomson sphere is very smeared out. Therefore, a detector such as \textit{WISPR} would not    be sensitive to electrons that are concentrated near the Thomson sphere, but rather to a much broader region on either side of the Thomson sphere ($\approx\chi_{TS} \pm 45^\circ$) that is called the Thomson plateau \citep[TP,][]{Howard2012}. An illustration of this effect for \name{WISPR} is given in \citet[][figure 14]{Poirier2020}. \\

Although there already are several numerical implementations of this theory within the scientific community \citep[e.g. in the \software{FORWARD} tool:][]{Gibson2016}, we opted to develop a new algorithm that we can tailor to the specific constraints of \name{WISPR} and the needs of this study. The procedure is summarised below.\\

\emph{Instrument definition: }Given both ephemeris (positioning) and pointing information for our virtual instrument, we build a 2D matrix of LOS coordinates.

\emph{Grid optimisation}: A dynamic grid refinement algorithm adjusts the sampling along each LOS, so as to capture at best the smallest physical structures in the simulation box (see Appendix \ref{sec:numerical}). The sample points are distributed from \textit{PSP}, and pass through and beyond the Thomson \Nico{sphere}. For the \name{WISPR} images synthesised in this work, this represents $241$ million   sample points to be optimised.

\emph{Thomson scattering computation}: Given a 3D simulated datacube of electron density, the Thomson scattering formula (\Nico{Eq. }\ref{eq:Thomson_scattering}) is computed at each sample point. This includes beforehand an interpolation step that can be very costly, given the large number of sample points and the size of the input datacubes used in this work ($(n_r,n_\theta,n_\phi)=(768,384,360)$ and $(256,160,320)$ for the idealistic and  fully fledged 3D set-up, respectively).

\emph{LOS integration}: The synthetic image is finally obtained by summing up all local contributions to the total brightness along each LOS.\\

\textit{WISPR} is a detector placed on a rapidly moving observatory sweeping extended regions of the solar corona in only a few days, together with a rapid variation in its distance to the Sun. The \textit{WISPR} FOV must therefore be updated very regularly, which is done by rerunning phase 1 and 2   for every single image to be synthesised. As \name{WISPR} is also much closer to the imaged coronal structures, it is critical to keep an accurate tracking of \textit{WISPR}'s pointing by using  the World Coordinate System (WCS). For this purpose,  phase 1 exploits the \software{IDL} routines provided by the \name{WISPR} instrument team through the \software{SolarSoft} library. \\

For the idealistic dipolar numerical set-up, however, we opted for a simple user-defined FOV that we can easily control to test different scenarios  in a sandbox. This allows us to simulate various viewing conditions that \name{WISPR-I} have encountered (or may in the future) at distinct phases of the solar cycle. We define a FOV representative of \name{WISPR-I} in the helioprojective--Cartesian frame with $HPLN=(10,50)^\circ$(azimuthal angle) and $HPLT=(-20,20)^\circ$(elevation angle) where $(HPLN=0,HPLT=0)$ points towards solar centre. We assume a null roll angle for simplicity. The helioprojective frame is a sphere centred at the observer position, which needs to be defined as well. We assumed a \name{PSP}-Sun distance of $35\ \rm{R_\odot}$ and $10\ \rm{R_\odot}$; the first is an average between the March 2022 and April 2021 events presented earlier, and the second  is intended to represent the closest approach that \name{PSP} will ever reach in 2024. The remaining parameter is the latitude of our virtual observer; the longitude does not matter since the idealistic simulation is axisymmetric about the solar rotation axis. Varying the latitude $\theta$ allows us to mimic different inclinations of the streamer belt from the  \name{WISPR} perspective, where large (or small) $\theta$ (in absolute value) are intended to be representative of a streamer belt seen face-on (or edge-on)  during solar maximum (or minimum) conditions. An inclination angle $\theta$ of $0^\circ$ and $\minus40^\circ$ has been assumed, for comparison with the April 2021 and March 2022 event, respectively.

\section{Modelling: Results}
\label{sec:results}

\subsection{Idealistic simulation of a dipolar corona}
\label{subsec:results_2D}

\begin{figure*}[]
\centering
\includegraphics[width=0.7\textwidth]{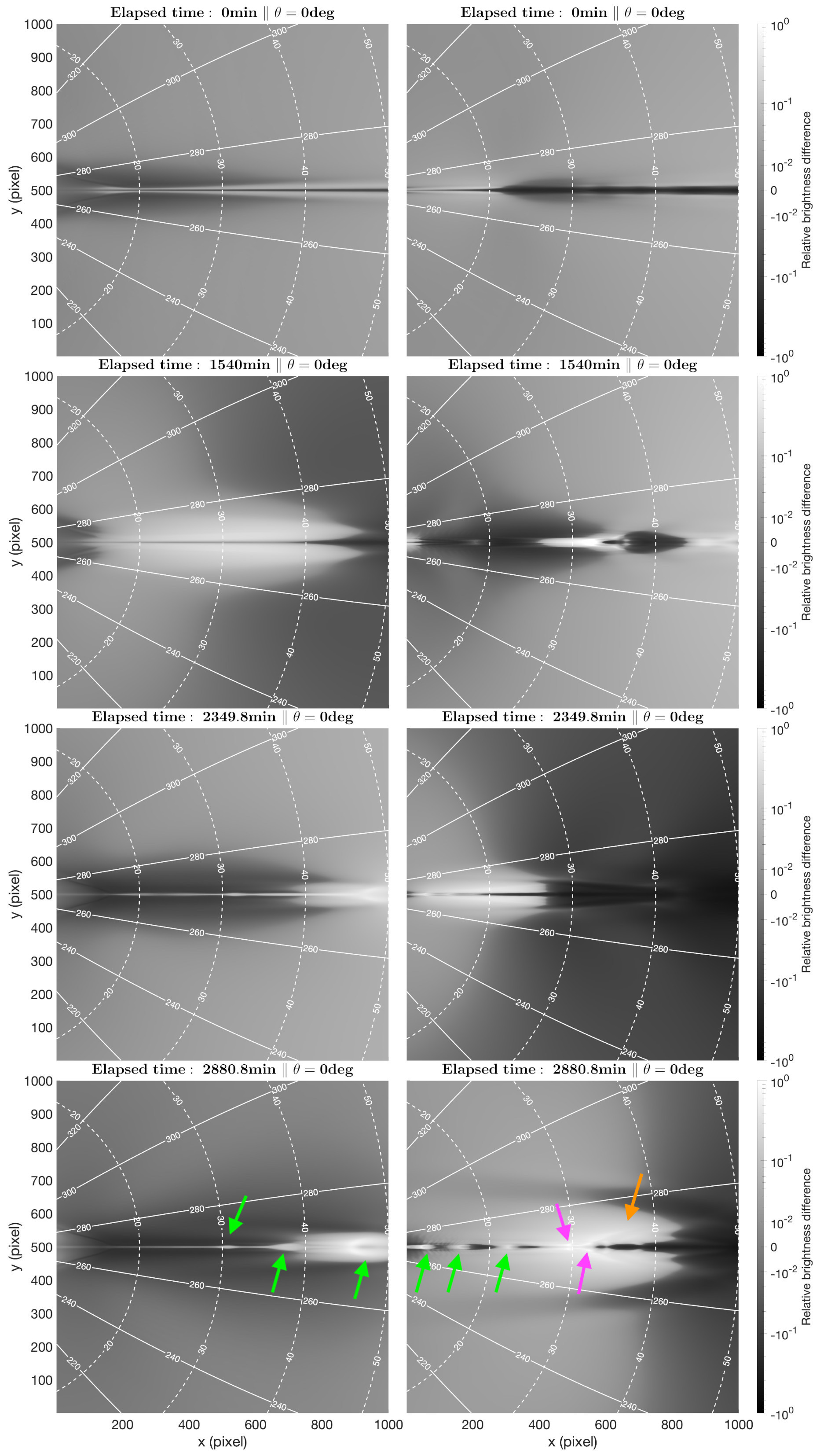}
\caption[]{ Synthetic \name{WISPR-I} images. The images were   produced from the idealistic dipolar set-up introduced in Sect. \ref{subsec:model_2D}, using a  layout similar to that in  Fig. \ref{fig:PLUTO2D}, assuming that \name{PSP} is located at $10\ \rm{R_\odot}$ (left column) and $35\ \rm{R_\odot}$ (right column). \Nico{This figure depicts the case of a flat streamer seen} edge-on by \name{WISPR} ($\theta=0^\circ$). \Nico{The relative brightness difference with the mean brightness computed over the full interval is colour plotted} using a symmetrical logarithmic scale (with linear scale below a cut-off value of $10^{-2}$). Isolines of the position ($\psi$) and elongation ($\epsilon$) angles (in degrees) are also plotted as solid and dashed white lines respectively. \Nico{The coloured arrows point to transient
structures (see main text for discussion).} An animated version of this figure is available online at \url{https://doi.org/10.5281/zenodo.8135596}.
\label{fig:SYNTH_IMG_PLUTO2Da}}
\end{figure*}

\begin{figure*}[]
\centering
\includegraphics[width=0.7\textwidth]{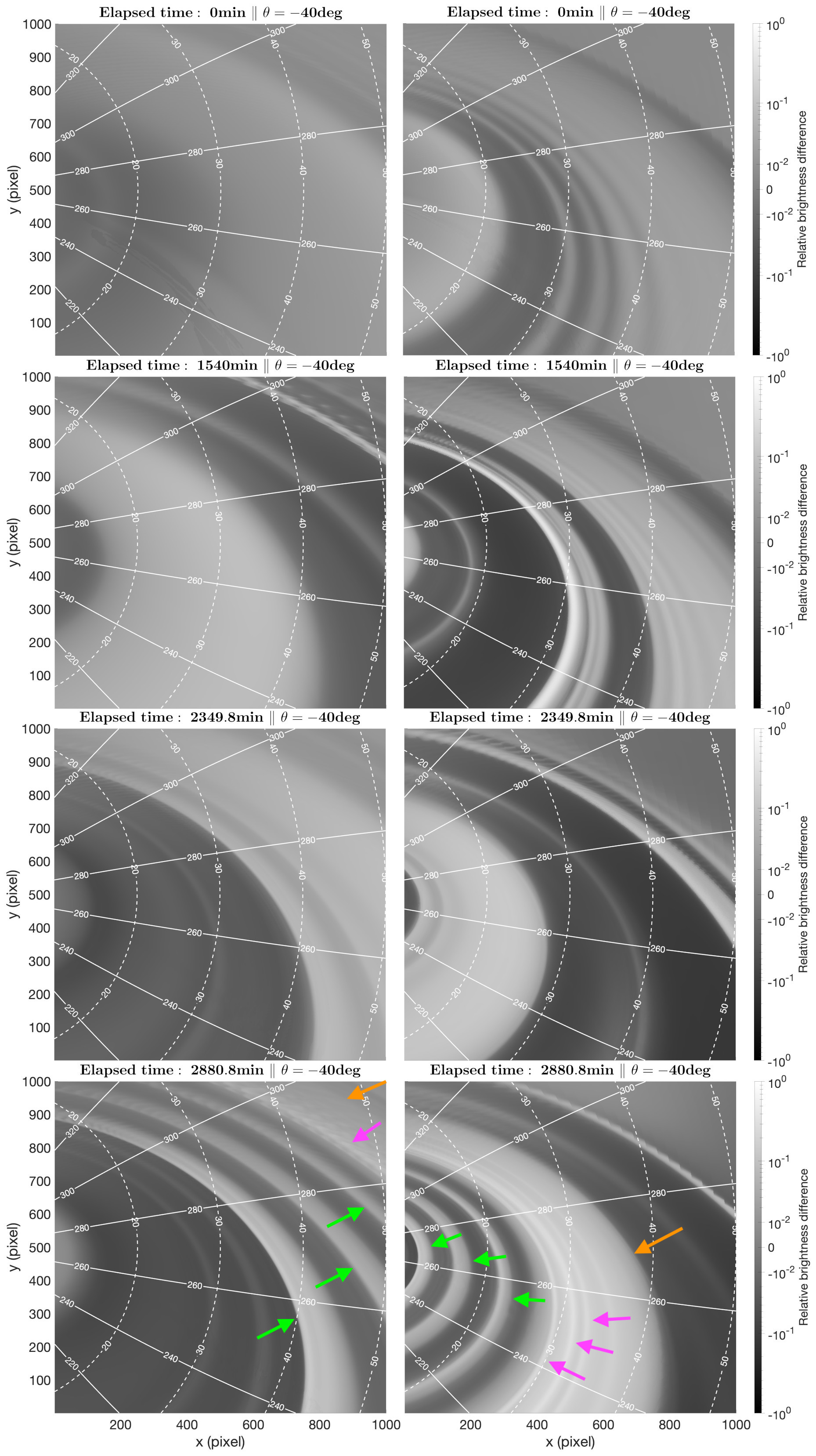}
\caption[]{ Same as Fig. \ref{fig:SYNTH_IMG_PLUTO2Da}, but for   a flat streamer seen at a $\theta=\minus40^\circ$ inclination angle by \name{WISPR}. An animated version of this figure is available online at \url{https://doi.org/10.5281/zenodo.8135596}.
\label{fig:SYNTH_IMG_PLUTO2Db}}
\end{figure*}

In Appendix \ref{appendix_absB} we gather the raw (absolute brightness) synthetic \name{WISPR} images produced from the idealistic dipolar modelling set-up introduced in Sect. \ref{subsec:model_2D}. To enhance the visibility of transient structures, we follow the base difference method where a background image (here computed as the average brightness over the entire time interval) is subtracted from each individual image. The resulting base-difference synthetic images are shown in Figs. \ref{fig:SYNTH_IMG_PLUTO2Da}-\ref{fig:SYNTH_IMG_PLUTO2Db}, where bright or dark colours respectively correspond to   an enhancement or depletion in electron density with respect to the background solar wind. These base-difference images reveal faint brightness variations much more clearly, and thanks to the new adaptive grid refinement method developed for this paper, small-scale density structures are rendered with great precision. This manifests as very smooth brightness variations across the LOS, where otherwise sharpness would   indicate an inappropriate sampling along the LOS. Some small spurious features (especially in the $\theta=\minus40^\circ$ case) are sometimes seen. These are remnant artefacts from the adaptive grid refinement method that   need further adjustments (discussed in Sect. \ref{sec:discussion}). \\

\noindent \textit{\textbf{WL signatures}}\\
We focus first on the two right-hand side panels of \Nico{Figs. \ref{fig:SYNTH_IMG_PLUTO2Da}-\ref{fig:SYNTH_IMG_PLUTO2Db}}, where the distance of \name{PSP} is taken close to that of the April 2021 and March 2022 events observed by \name{WISPR} (i.e. $35\ \rm{R_\odot}$). The four rows cover two full cycles of the development of the tearing instability. The $t=2880.8\ \rm{min}$ snapshot (bottom rows) is   the one that illustrates   the different phases best. We can group the synthetic WL signatures in two main families:    bright diffuse patches   and    bright and more concentrated emissions.  They are described below. \\

The bright diffuse patches  result from the main onsets of the tearing instability (i.e.  the ballooning mode; see the orange arrows). Due to their rather large scale, they are likely the \Nico{Sheeley blobs} that have long been observed from 1 AU, and first detected by \name{SoHO-LASCO} \citep{Sheeley1997}. When seen edge-on ($\theta=0^\circ$, Fig. \ref{fig:SYNTH_IMG_PLUTO2Da}) they show quite significant brightness enhancement of up to $\approx 35\%$; instead,  when seen with some inclination ($\theta=\minus40^\circ$, Fig. \ref{fig:SYNTH_IMG_PLUTO2Db}) they appear slightly dimmer with brightness enhancements below $\approx 10\%$. Because the simulated transients have here an infinite extent in azimuth, they show drifting signatures towards the FOV edges as they pass over \name{WISPR} location (see second row of Fig. \ref{fig:SYNTH_IMG_PLUTO2Da}, right panel), which are  similar to WL signatures that were observed when \name{WISPR} approached and went through the streamer belt (\citealt{Howard2019,Poirier2020}; see also the simulations by \citealt{Liewer2019}). Nonetheless, having infinite azimuthal extents for such transients is not realistic, as clearly shown by the March 2022 event, and we   show in Sect. \ref{subsec:results_3D} that this can be solved using the fully fledged 3D modelling set-up.

The bright and more concentrated emissions exhibit quasi-periodic formation (see the green and purple arrows). Similar spatial distributions and widths to those of   the April 2021 and March 2022 events observed by \name{WISPR} can be seen in the lower right panels of Figs. \ref{fig:SYNTH_IMG_PLUTO2Da} and \ref{fig:SYNTH_IMG_PLUTO2Db}, respectively. These quasi-periodic structures develop at the back of the main onsets described above, and can be connected to the long-observed hourly periodicities measured both remotely and in situ in the slow solar wind \citep{Viall2010,Viall2015,Kepko2016}. These structures are smaller in size than the streamer blobs discussed   above, and hence we   expect them to contribute less to the total brightness integrated along the LOS. Conversely, they exhibit much higher brightness enhancements because they contain a much higher concentration of plasma. Their brightness variation ranges $\approx 20-100\%$ (edge-on case, $\theta=0^\circ$, Fig. \ref{fig:SYNTH_IMG_PLUTO2Da}) and $\approx 1-40\%$ (face-on case, $\theta=\minus40^\circ$, Fig. \ref{fig:SYNTH_IMG_PLUTO2Db}). Some of these structures propagate faster, and as a result   sometimes   coalesce with their preceding fellows or even merge with the main onset (see the purple arrows). \\

In terms of brightness variation, there is a fair agreement between the simulated quasi-periodic structures and the transients observed by \name{WISPR}, that is $\approx 80-95\%$ for the April 2021 event (edge-on case, to be compared with $\theta=0^\circ$) and $\approx 30-50\%$ for the March 2022 event (face-on case, to be compared with $\theta=\minus40^\circ$). On the other hand, the simulated ballooning modes (which can be associated with the Sheeley blobs) may be more difficult to see from the \name{WISPR} perspective as their signature is fainter and more diffuse. They are also less likely to be detected by \name{WISPR} due to their long periodicity, as we   describe below  in  `\emph{Periodicities}'. 

Access to shorter heliocentric distances might help to better resolve both structures as illustrated in the left panels of Figs. \ref{fig:SYNTH_IMG_PLUTO2Da}-\ref{fig:SYNTH_IMG_PLUTO2Db}, assuming the hypothetical $\simeq 10\ \rm{R_\odot}$ to be reached by \name{PSP} in 2024 at the closest approach. There the shrinking of the Thomson plateau (i.e. of the sensitive area of \name{WISPR}; see Sect. \ref{subsec:synthetic}) should allow transients within streamers to more easily stand out from the background emissions. \name{WISPR} will also be able to observe these structures right in their formation region ($3-7\ \rm{R_\odot}$), and therefore might provide new clues about the tearing instability occurring at the HCS. \\

\begin{figure*}[]
\centering
\includegraphics[width=0.98\textwidth]{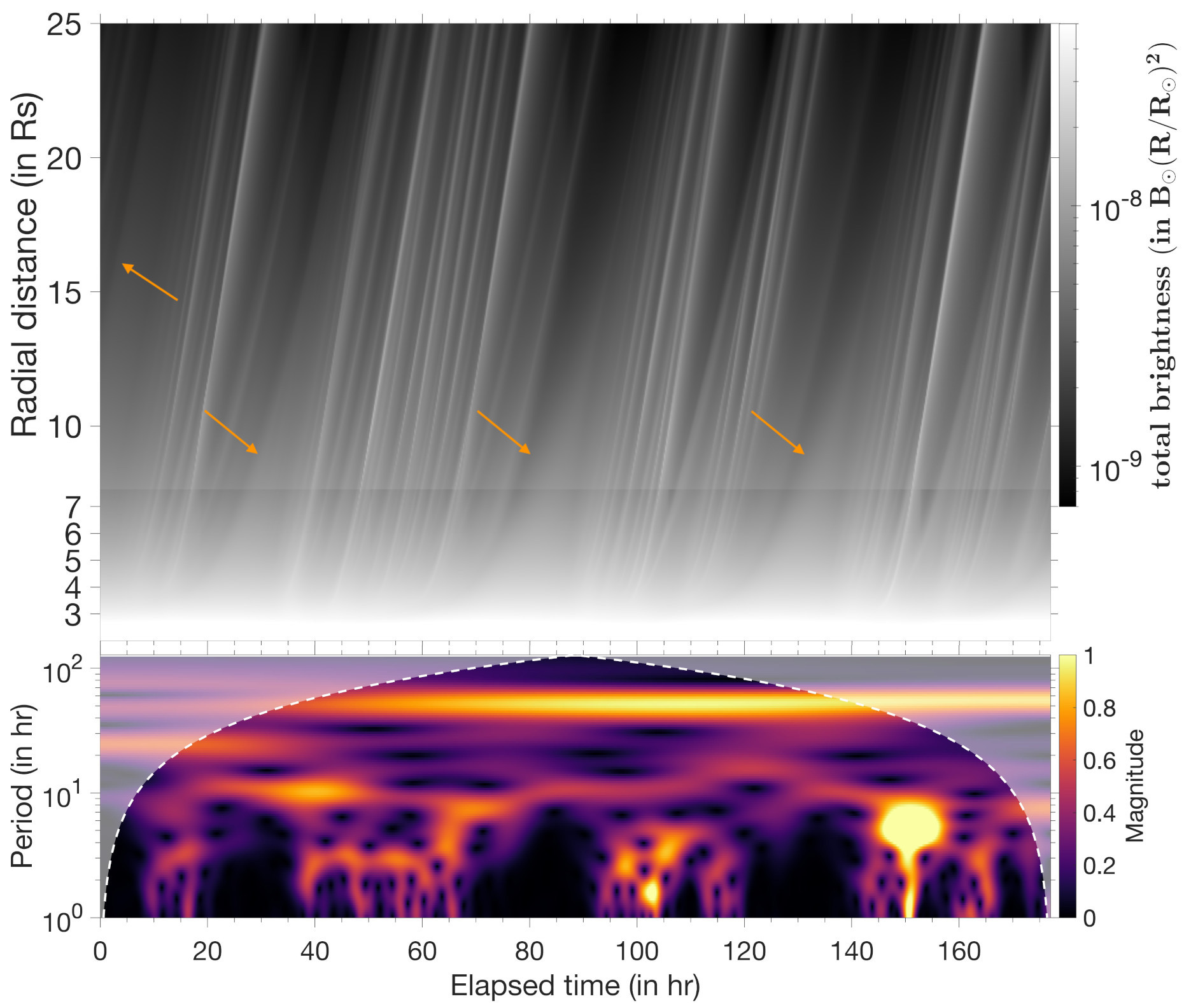}
\caption{ Time analysis of  WL transients simulated from the idealistic dipolar set-up introduced in Sect. \ref{subsec:model_2D}. Top panel: Synthetic \name{WISPR-I} J-maps built from the two-image sequences shown in Fig. \ref{fig:SYNTH_IMG_PLUTO2Da} and extracted along the solar equator ($\psi=270^\circ$). Bottom panel: Associated period decomposition at $r=10\ \rm{R_\odot}$ using a 1D continuous wavelet transform \citep[Morse wavelet,][]{Torrence1998}.
\label{fig:SYNTH_JMAPS_PLUTO2Da}}
\end{figure*}

\noindent \textit{\textbf{Propagation velocities and acceleration profiles}}\\
We now look at the kinematics of the simulated transients, by making J-maps as those shown in Fig. \ref{fig:WISPR_Jmaps}. 

A synthetic J-map for the $\theta=0^\circ$ case is given in Fig. \ref{fig:SYNTH_JMAPS_PLUTO2Da}, where the slit has been taken at the solar equator (i.e. along the streamer). The $\theta=\minus40^\circ$ case is not shown because it does not  change   the rest of the analysis much. We retrieve here the two main families of WL signatures identified in the previous section. First, the wide patches associated with the main onsets of the tearing instability (i.e. the ballooning mode; see   orange arrows). Second, the more concentrated emissions associated with quasi-periodic transients, not pinpointed here as they clearly stand out as bright thin stripes from the rest. 

The ballooning modes show quite different signatures in the J-map, with more curvature and less inclination. Their lower inclination indicates that they propagate at a slightly lower speed than the quasi-periodic structures, which are then likely to merge together as discussed previously. Their curvature may also be indicative of a more progressive acceleration until they reach their terminal speed after $\approx 10-15\ \rm{R_\odot}$; in contrast, the fast quasi-periodic structures show clear constant speed profiles. Here the acceleration patterns are likely to be actual and not apparent accelerations as our \name{WISPR-I} FOV remains static in this simulation set-up (see Sect. \ref{subsec:synthetic}), although LOS-integration effects could still contribute to these curvatures, as already discussed in Sect. \ref{subsec:WISPR_obs_Jmaps}. In both cases, the simulated transients reach a terminal speed of $\approx 250\ \rm{km/s}$, which corresponds to the bulk speed of the very slow and dense wind at the core of the streamer belt in the simulation.\\

\noindent \textit{\textbf{Periodicities}}\\
The bottom panel of Fig. \ref{fig:SYNTH_JMAPS_PLUTO2Da} shows the dominant periodicities over four full cycles of the tearing instability. The time elapsed  between each main onset of the tearing instability (i.e. the ballooning modes) is quite variable, reaching $\approx 25\ \rm{hr}$ between the first two and $\approx 50\ \rm{hr}$ for the others. Here, this  is  mostly driven by how fast helmet streamer loops can grow due to pressure imbalance, and hence is highly sensitive to local coronal conditions. This long periodicity is then likely to vary significantly from one simulation to another, and more generally over solar longitudes and along the solar cycle (see the discussion in Sect. \ref{subsec:context_1au}). In comparison, a shorter $\approx 8-16\ \rm{hr}$ period had been detected from past \name{STEREO-A} observations \citep{Sanchez-Diaz2017b}, but this study focused on a few specific events around the solar maximum of cycle 24. More recently, an analysis of the observations taken by the \name{STEREO-A COR-2} coronagraph near solar minimum has shown density variations in the streamer belt on timescales of 0.5-2 days \citep{Morgan2021}, which this time agree with our simulation. 
Due to a fast and highly elliptical orbit, \name{WISPR} is not appropriate to detect such long periodicities. Therefore, the legacy 1 AU observatories remain valuable assets, which in complement to the recently launched \name{Solar Orbiter} might finally allow us to better parametrise such events.

Regarding the quasi-periodic structures generated  between each main onset, a wavelet power spectrum \citep{Torrence1998} reveals dominant periods around $\approx 2-3\ \rm{hr}=120-180\ \rm{min}$ and $\approx 7-10\ \rm{hr}$. In addition, we   note  that the simulation   exhibits some periodicities as low as $\approx 1.5\ \rm{hr}=90\ \rm{min}$ (at $t=100-105\ \rm{hr}$). These results are in good agreement with the $\approx 90-180\ \rm{min}$ to $\approx 8-16\ \rm{hr}$ periods that have been typically detected from 1 AU \citep{Viall2010,Viall2015,Kepko2016,Sanchez-Diaz2017a}. More specifically, they also agree with those measured during the April 2021 ($130-175\ \rm{min}$) and March 2022 ($110-120\ \rm{min}$) events observed by \name{WISPR}.

\subsection{Case study simulation of the ninth \name{PSP} encounter}
\label{subsec:results_3D}

We now exploit the fully fledged 3D simulation set-up introduced in Sect. \ref{subsec:model_3D} and detailed in \cite{Reville2022}. Starting from April 2021 (eighth encounter), interpreting \name{WISPR} observations has become highly challenging even with the use of such state-of-the-art modelling. This is primarily due to \name{PSP} diving much deeper inside the nascent solar wind and to an increase in the solar activity resulting in a much more structured corona. Tremendous efforts in tuning-up the model parameters would be required for a fair one-to-one comparison with the actual \name{WISPR} observations taken during the ninth encounter (August 2021). A dynamic update of the magnetic map at the inner boundary would also be essential to reach this goal. We leave this  for future works, but there is still a valuable set of information that we can extract from this simulation to feed the current discussion.

\begin{figure}[]
\centering
\includegraphics[width=0.5\textwidth]{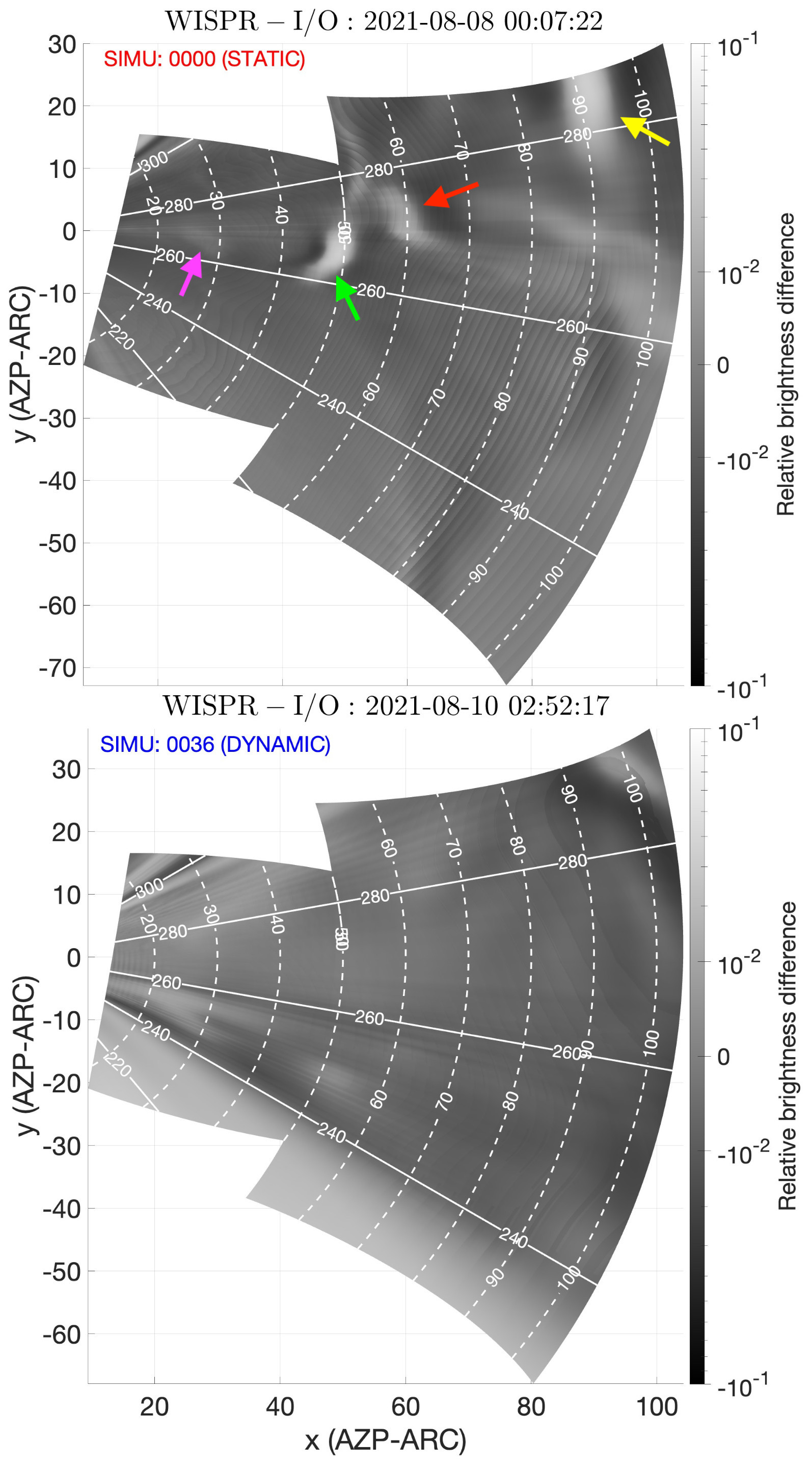}
\caption[]{ Synthetic images of  inner \name{WISPR-I} and outer \name{WISPR-O} telescopes combined. The images were  produced with the fully fledged 3D simulation set-up introduced in Sect. \ref{subsec:model_3D}. A mean background was subtracted from each image using a sliding temporal window ranging $\approx 5-10\ \rm{hours}$ (running difference method). The relative brightness difference to this mean background is colour plotted with a symmetrical logarithmic scale (with linear scale below a cut-off value of $10^{-2}$). As the simulation is $\approx 22\ \rm{hours}$ long, only a short temporal window around perihelion could be treated in a self-consistent manner (dynamic mode). Outside this dynamic interval, images were then   produced with either the first or last simulation snapshot (static mode). An animated version of this figure is available online at \url{https://doi.org/10.5281/zenodo.8135596}.
\label{fig:SYNTH_IMG_PLUTO3D}}
\end{figure}

We present in figure \ref{fig:SYNTH_IMG_PLUTO3D} the result of our forward modelling method applied to this simulation. Similarly to Figs. \ref{fig:SYNTH_IMG_PLUTO2Da}-\ref{fig:SYNTH_IMG_PLUTO2Db}, a difference method is used to better visualise brightness fluctuations due to transient propagating structures. However,  computing a mean background over the entire time interval is no longer appropriate  since our virtual \name{WISPR} observer is no longer static. We then follow the well-known running difference method here, where the mean background image is computed over a sliding temporal window. For completeness and a more realistic impression, the raw synthetic products in absolute brightness are also shown in Fig. \ref{fig:SYNTH_IMG_PLUTO3D_abs}. \\

In contrast to the idealistic modelling set-up, a wealth of signatures are produced here with a rich diversity of shapes, extents, and locations. They are indicated in Fig. \ref{fig:SYNTH_IMG_PLUTO3D} by coloured arrows, which can be directly connected to the flux rope structures present in the simulation snapshot shown in Fig. \ref{fig:PLUTO3D}. Transients can be seen throughout the \name{WISPR-I/O} FOVs. Overall they show great topological similarity with the April 2021 and March 2022 events observed by \name{WISPR-I} (see Fig. \ref{fig:WISPR_blobs}), namely arch-like (top panel) and blob-like (bottom panel) signatures representative of flux ropes seen face-on and edge-on, respectively. This time, the fully fledged 3D set-up is more realistic with arch signatures of finite (and not infinite) spatial extent because it includes secondary (or pseudo) streamer structures that were missing in the idealistic dipolar set-up (see Sect. \ref{subsec:model_3D}). 

Furthermore, using Fig. \ref{fig:SYNTH_IMG_PLUTO3D} we can estimate that the simulated transients induce a similar increase in relative brightness of $\approx1-4\%$ during their passage, although they are located at different distances from \name{WISPR}. These values are quite low compared to the brightness variations   estimated above from \name{WISPR} observations and the dipolar corona set-up  because the background field  here is computed over a much shorter temporal window, which is not   representative of the emissions from the background streamers. For this purpose we should use instead the raw synthetic images (in absolute brightness)  shown in Fig. \ref{fig:SYNTH_IMG_PLUTO3D_abs}, and compare the absolute brightness inside and just outside the transients (similarly to what we did for the real \name{WISPR} observations). By doing so we obtain more reasonable brightness variations of $\approx8-17\%$. \\

Many of the brightness variations visible in Fig. \ref{fig:SYNTH_IMG_PLUTO3D} do not result from a propagating transient feature, but from a change in the viewing conditions of \name{WISPR} as \name{PSP} flies rapidly along its orbit. This has been seen many times,  for instance from the drift of streamer rays towards the FOV edges \citep{Liewer2019,Howard2019,Poirier2020}. This effect can be visualised in a supplementary movie provided with Fig. \ref{fig:SYNTH_IMG_PLUTO3D}. Starting from  August 09, 2021, 18:37 UT, the movie shows the effect of \name{PSP} moving throughout the streamer belt, for the first time with a fully dynamic 3D modelling set-up that  includes the self-generation of streamer transient structures. As \name{WISPR}   moves closer to (and probably   into) these imaged transients, their morphology changes quite significantly, which is  the consequence of a change in the perspective and of a change in the sensitivity area of \name{WISPR} (see Sect. \ref{subsec:synthetic}). Our comparison basis with actual observations can only be qualitative here for all the reasons mentioned above, such as the fact that the simulation does not cover the entire \name{WISPR} interval studied here (see Sect. \ref{subsec:model_3D}). Even so,  we were able to identify many similarities between this synthetic movie and the newest \name{WISPR} observations starting from August 2021 (official movies for all \name{WISPR} observations can be found online\footnote{\url{https://wispr.nrl.navy.mil/encounter-summaries}}).

\section{Discussion}
\label{sec:discussion}

In this study we focused only on a few transient events observed by \name{WISPR} that we considered as promising candidates produced by the streamer pinch-off reconnection mechanism. The April 26 and March 01 events were picked especially  for their great visibility, but do not represent the full set of observations. \name{WISPR} observations show a plethora of transient features, in particular in the late \name{PSP} encounters (from September 2020 and on). Further studies are required to infer those transient properties in a statistical manner \citep[see e.g. the upcoming paper from][]{Viall2023}.

Although the two modelling set-ups presented in this work show a great potential for interpreting some of the transient nature of \name{WISPR} images, there are still many  improvements  that are needed. Significant advances have been made in the forward modelling procedure compared to our previous study \citep{Poirier2020}, where most of the previous difficulties have been resolved. The adaptive grid refinement method allows for a much more accurate sampling of the LOS, which results in a very smooth rendering of small-scale density structures. Despite the high precision achieved, we still note some remaining artefacts of minor importance in the difference-based synthetic products. For the sake of computational tractability, for each individual image we had to keep the number of optimisation steps for the LOS sampling to a maximum of 30 in our case. This implies that the optimisation procedure may not fully converge for all LOS, at times generating some spurious features.

These are minor inconveniences, and now the performance of our white-light rendering code is primarily limited by the quality of the input simulation. Future efforts should then concentrate on pushing existing solar coronal or wind models beyond their current capabilities in order to allow for a thorough understanding of the latest \name{WISPR} observations. In light of this work, we present   below \Nico{some areas} of progress that could be addressed. \\

\emph{Coronal structure fidelity: }Current comparisons against \name{WISPR} observations greatly suffer from a misplacement of the main streamer belt and pseudo-streamer structures. This highlights the unique capability of \name{WISPR} to provide more stringent constraints  to current coronal models. For instance the magnetic map set at the inner boundary is known to have a critical impact on the performance of existing MHD models. Having a magnetic map that is updated dynamically over time will be necessary improve comparisons with recent \name{WISPR} observations. On-going efforts have also been carried on towards a systematic benchmarking of MHD models against observations \citep[see e.g.][]{Badman2022}, and \name{WISPR} observations could both greatly benefit from and support such works.\\

\emph{Spatial resolution: }Although the fully fledged 3D simulation set-up depicts a much more realistic solar coronal structure,  it only permits a partial development of the tearing instability compared to the idealistic but much more (spatially) resolved 2.5D set-up. Future works will have to refine the spatial resolution around the HCS further, but that is extremely challenging in such global 3D MHD models while maintaining computational tractability. \\
 
\emph{Temporal cadence and duration: }Together with a sufficiently spatially resolved model, temporal cadence is also important to track these streamer transients as they rapidly propagate throughout the \name{WISPR} FOV. This will be important especially in the years to come when \name{PSP} will be able to sample these transients right in their formation region (i.e. the $10\ \rm{R_\odot}$ case treated with the 2.5D set-up). Having simulation snapshots at a higher cadence than \name{WISPR} could also help to get a better match by allowing the synthetic images to mimic the `blurring' effect of exposure time. Furthermore, simulation duration of more than one day would be preferable to maximise the scientific output. This can easily be achieved in a 2.5D context, but implies large amounts of data in full 3D modelling set-ups, a challenge to reach with modern computational facilities.

\section{Conclusion}
\label{sec:conclusion}

The variability of the slow wind that originates from streamers has been analysed in light of the latest observations taken by \name{WISPR}. A few transient events have been identified with periodicities that are consistent with the previous $90-180\ \rm{min}$ range detected from near 1 AU observations. The pinch-off reconnection mechanism occurring at the tip of helmet streamers has long been predicted as a potential source mechanism of these quasi-periodic structures. 

For the first time this work provides strong evidence  to support this scenario using two advanced MHD models of the solar wind and corona, each  with its own pros and cons. Both give rise to the same fundamental process, however. First there is \Nico{a pressure} instability of the coronal loops that are lodged beneath the helmet streamers that allows them to rise in the corona. They stretch to a point where the current sheet that develops at their back becomes so thin that magnetic reconnection eventually occurs via the tearing instability. A large flux rope made of streamer material is then released (i.e. the main onset or ballooning mode), corresponding to the \Nico{Sheeley} blobs that were first detected in \name{SoHO-LASCO}. Behind this main ejecta follows the further development of the tearing instability at the HCS, that generates a myriad of quasi-periodic smaller-scale structures. We \Nico{show} that these quasi-periodic structures exhibit local density enhancements that are strong enough to be detected by \name{WISPR}, and that they actually show great topological similarities with two real events captured by \name{WISPR}. In addition, the simulated quasi-periodic structures have periodicities that agree well with these events, and also more generally with the $90-180\ \rm{min}$ range detected in past observations. 

These quasi-periodic structures could be reproduced in an idealistic dipolar set-up thanks to a very high spatial resolution at the HCS. However, this set-up lacked some consistency to simulate properly the actual WL signatures observed by \name{WISPR}. A global fully fledged 3D MHD model was then necessary to simulate the aspect of these structures in a self-consistent manner. However,  the coarser spatial resolution did not allow   the full development of the tearing instability, and thus of the quasi-periodic structures where only \Nico{the $\gtrsim 4\ \rm{hours}$ long periodicities could be reproduced}. \\

This work highlights the importance of the tearing instability occurring at the tip of streamers  to fuel the long-observed high variability of the slow solar wind. Furthermore, we discuss   how extremely challenging the latest (and upcoming) \name{WISPR} observations are to interpret, even just for the quasi-steady component of the slow wind because \name{PSP} is diving deeper and deeper within a solar corona that becomes highly structured with the rising phase of the solar cycle. Therefore, \name{WISPR} offers new stringent constraints to push existing models of the solar corona and wind beyond their current capability, which in turn should help in better understanding \name{WISPR} observations.

\section*{Acknowledgements}
The authors are indebted to an anonymous referee whose valuable suggestions greatly improved this work.

This research has been funded by the ERC SLOW\_SOURCE (DLV-819189) and NRC ORCS (324523) projects. A.K. was supported by NASA’s Parker Solar Probe mission under contract NNN06AA01C.

The authors are grateful to Nicholeen Viall and Angelos Vourlidas for insightful discussions, and the \name{WISPR} team for providing the data. The \name{Wide-Field Imager for Parker Solar Probe} (\name{WISPR}) instrument was designed, built, and is now operated by the US Naval Research Laboratory in collaboration with Johns Hopkins University/Applied Physics Laboratory, California Institute of Technology/Jet Propulsion Laboratory, University of Gottingen, Germany, Centre Spatial de Liege, Belgium and University of Toulouse/Research Institute in Astrophysics and Planetology. \name{Parker Solar Probe} was designed, built, and is now operated by the Johns Hopkins Applied Physics Laboratory as part of NASA's Living with a Star (LWS) program.

The authors also thanks A. Mignone and the PLUTO development team, on which WindPredict-AW is based. The 2.5D and 3D WindPredict-AW simulations were performed on the Jean-Zay supercomputer (IDRIS), through the GENCI HPC allocation grant A0130410293.

The photospheric magnetic maps used in this work are produced collaboratively by AFRL/ADAPT and NSO/NISP. The SOHO/LASCO data are produced by a consortium of the Naval Research Laboratory (USA), Max-Planck-Institut für Aeronomie (Germany), Laboratoire d’Astronomie (France), and the University of Birmingham (UK). \name{SOHO} is a project of international cooperation between ESA and NASA.

This work made use of the data mining tools \software{AMDA}\footnote{\url{http://amda.irap.omp.eu/}} developed by the Centre de Données de la Physique des Plasmas (CDPP) and with financial support from the Centre National des Études Spatiales (CNES). This study also used the NASA Astrophysics Data System (ADS\footnote{\url{https://ui.adsabs.harvard.edu/}}), the open-source GNU Image Manipulation Program (\software{GIMP}\footnote{\url{https://www.gimp.org/}}) and the \software{ImageJ}\footnote{\url{http://imagej.nih.gov/ij}} image processing tool developed by Wayne Rasband and contributors at the National Institutes of Health, USA.

\bibliography{biblio}
\bibliographystyle{aa}

\begin{appendix}
\section{Computational challenges and method}
\label{sec:numerical}
Having underresolved LOSs has been shown to have significant consequences on the synthetic images produced \citep[e.g. artefacts, missing density structures; see][]{Poirier2020}, and as such $121$ points appeared to be sufficient to resolve the features studied in that previous work. However, the idealistic dipolar set-up introduced in Sect. \ref{subsec:model_2D} requires  resolving spatial structures as small as $0.1\ \rm{R_\odot}$. As a consequence, running again our synthesising script with $121$ points only was no longer adequate. A significant improvement compared to our previous work \citep{Poirier2020} has then been to push the LOSs resolution from $121$ to $241$ points, resulting in several computational challenges to tackle,  which are discussed below.

Synthesising a white-light image of $1000$-by-$1000$ pixels implies  computing multiple 3D matrices with up to $241$ million elements each, for LOSs resolved with $241$ points. In typical $32GB$ memory systems this quickly leads to a memory overflow. One main challenge has been to optimise the code so as to minimise the memory usage, and at the same time maximise the workload on CPUs. A prior step before actual computation is to break the image down into smaller sections, where the number of sections is adjusted automatically to ensure maximised performances. Each sub-section is then computed in parallel, hence using at most the capacity of current multi-core systems.

In theory the LOSs could be better resolved than $241$ points, resulting in more sub-sections to compute, and hence to a longer computational time. However, we restrained ourselves to $241$ points instead and worked on optimising the point distribution along each LOS. To do so one needs to have a prior idea of which portions of the LOSs need to be better resolved than others. The Thomson scattering (introduced in Sect. \ref{subsec:synthetic}) served as a basis to optimise the point distribution, following a two-step procedure described below. 

We start by defining a uniform grid in scattering angle $\chi$ that covers both the foreground ($\chi=90^\circ \rightarrow \chi=\chi_{max}$) and background ($\chi=\chi_{min} \rightarrow \chi=90^\circ$) with respect to the Thomson sphere ($\chi=90^\circ$), where $\chi_{min}=0^\circ$ and $\chi_{max}=180^\circ-\alpha$ are the asymptotic limits to the acceptable range of $\chi$ angles ($\alpha$ is the central angle between the LOS and the observer--Sun line). Using a $\chi$-defined uniform grid is convenient as it naturally produces a non-uniform grid in the path length $z$ (i.e. the distance along a LOS from the observer's position), with a minimum spatial step near the Thomson sphere. We make a first computation of the total brightness on this uniform grid. Before proceeding to the grid optimisation described afterwards, the spatial extent of each LOS is reduced to a region that accounts for most  of the total integrated brightness; we used  $99\%$ in this paper. The upper $\chi_u$ and lower $\chi_l$ limit to the integral of the total brightness (Eq. (\ref{eq:Thomson_scattering})) are determined when the ratio 
\begin{equation}
   \mathcal{R}=\frac{\sum_{\chi=90^\circ}^{\chi_{l,u}} n_e z^2 G dz}{\sum_{\chi=90^\circ}^{\chi_{min,max}} n_e z^2 G dz}
\end{equation}
reaches $0.99$ in both the foreground and the  background. This allows us to save more grid points in needed areas and to maximise the efficiency of the grid refinement step described in the next paragraph. \\

In the second step we   implemented an adaptive grid method following a similar approach to that used by \citet{Dorfi1987}, where the spatial refinement adapts dynamically according  to the physical structures to be resolved, which are  here the density structures along each LOS. For this purpose we define the grid point density by a function that includes the local distribution of the total (i.e. not polarised) WL intensity along each LOS
\begin{equation}
\begin{aligned}
    R^{k+1}_i &= \frac{1}{c}\sqrt{w_1\left(\frac{\sum_{i=1}^{i=nlos} \Delta\chi^k_i}{\chi_u-\chi_l}\right)^2 + w_2\left(\frac{I^k_{t,i}/\Delta z^k_i}{mean(I^k_{t,i}/\Delta z^k_i)}\right)^2} \\
    c &= \sqrt{w_1+w_2}
\end{aligned},
\end{equation}
where the lower $i$ and upper $k$ sub-scripts refer to the spatial \Nico{index (position along the LOS) and the optimisation iteration number}, respectively. This formulation allows us to apply multiple optimisation criteria pondered by their respective weights $w_*$, all chosen as equal to one in this paper (after performing several tests). The first criterion (left term) allows us to constrain the extent of the optimised grid near the previously determined $99\%$ range of interest. The second criterion purely depends on the studied physical system where more points are set where the local intensity is greater. One should make sure to use normalised quantities to get a proper balance between each criterion. Using the mean value of the integrated intensity $mean(I^k_{t,i}/\Delta z^k_i)$ along each LOS appears to work best. We note that $w_2=0$ would lead to a uniform grid in $\chi$. The actual spatial step in $\chi$ angle at the next \Nico{optimisation step} is then determined by
\begin{equation}
    \Delta\chi^{k+1}_i = \Delta\chi^k_i\frac{R^k_i}{R^{k+1}_i}
.\end{equation}
\Nico{This procedure is applied to each LOS and is repeated iteratively until the brightness integrated along each LOS converges to a stable value, which is defined as $<1\%$ variation with previous iteration in this work. To accelerate the optimisation process, the optimised grid is re-used from one image to another as a new initialisation instead of the $\chi$-defined uniform grid.}

\section{Raw synthetic products, absolute brightness}
\label{appendix_absB}
\begin{figure*}[]
\centering
\includegraphics[width=0.75\textwidth]{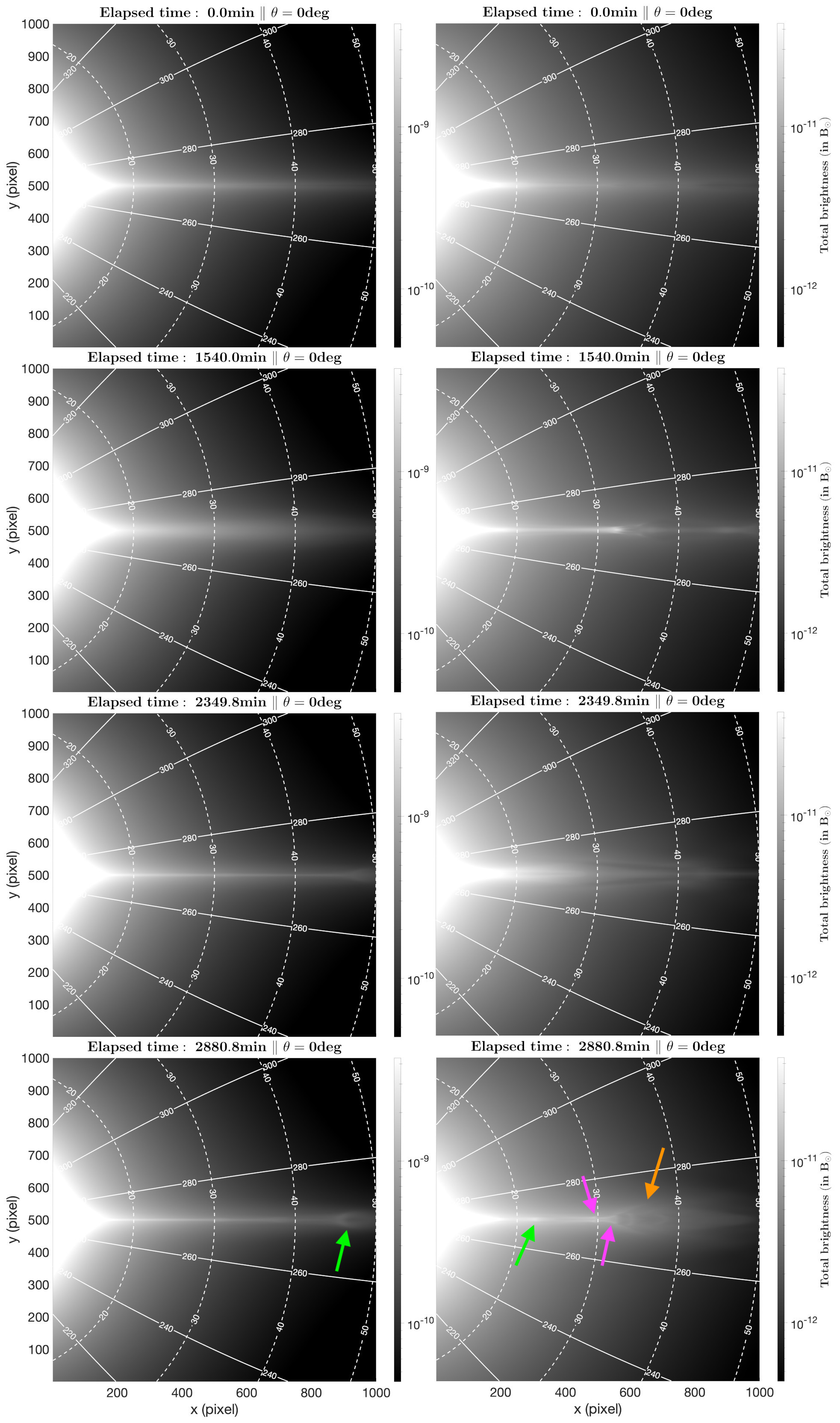}
\caption[]{ Same as Fig. \ref{fig:SYNTH_IMG_PLUTO2Da}, but for   absolute brightness, plotted in units of mean solar brightness $B_\odot=2.3e7\ \rm{W.m^{-2}.sr^{-1}}$ and in   logarithmic scale. An animated version of this figure is available online at \url{https://doi.org/10.5281/zenodo.8135596}.
\label{fig:SYNTH_IMG_PLUTO2Da_abs}}
\end{figure*}

\begin{figure*}[]
\centering
\includegraphics[width=0.75\textwidth]{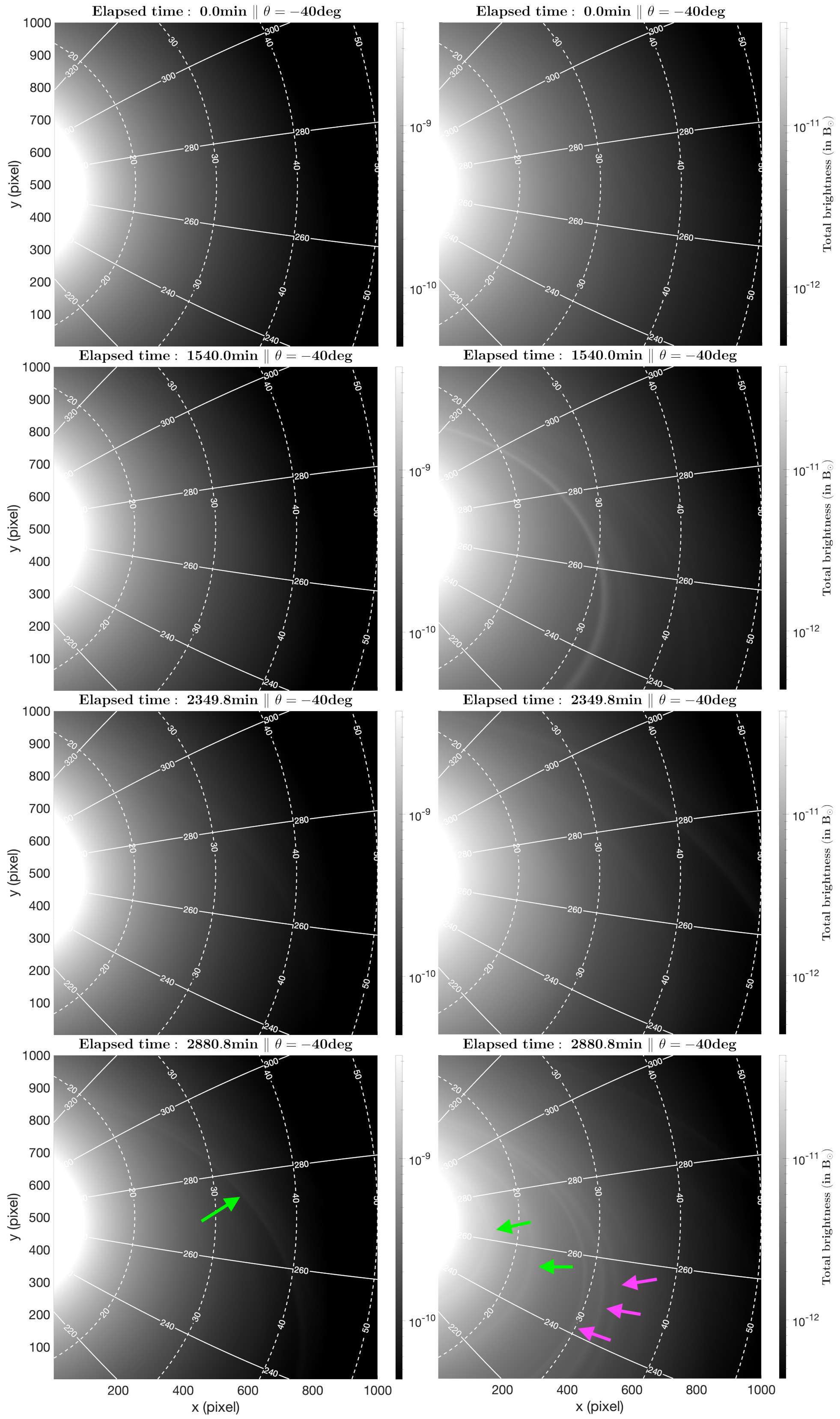}
\caption[]{ Same as Fig. \ref{fig:SYNTH_IMG_PLUTO2Db}, but for   absolute brightness, plotted in units of mean solar brightness $B_\odot=2.3e7\ \rm{W.m^{-2}.sr^{-1}}$ and in   logarithmic scale. An animated version of this figure is available online at \url{https://doi.org/10.5281/zenodo.8135596}.
\label{fig:SYNTH_IMG_PLUTO2Db_abs}}
\end{figure*}

In terms of absolute brightness most of the transient signatures remain relatively faint over the background solar wind (see the coloured arrows in Figs. \ref{fig:SYNTH_IMG_PLUTO2Da_abs}-\ref{fig:SYNTH_IMG_PLUTO2Db_abs}), even though our virtual \name{WISPR} observer is imaging them from a very close distance. The idealistic case of an inclined streamer belt ($\theta=\minus40^\circ$, Fig. \ref{fig:SYNTH_IMG_PLUTO2Db_abs}) is even dimmer compared to the LOS-aligned streamer belt case ($\theta=0^\circ$, Fig. \ref{fig:SYNTH_IMG_PLUTO2Da_abs}), as a much smaller portion of these transients is integrated along the LOS. 

\begin{figure*}[]
\centering
\includegraphics[width=0.7\textwidth]{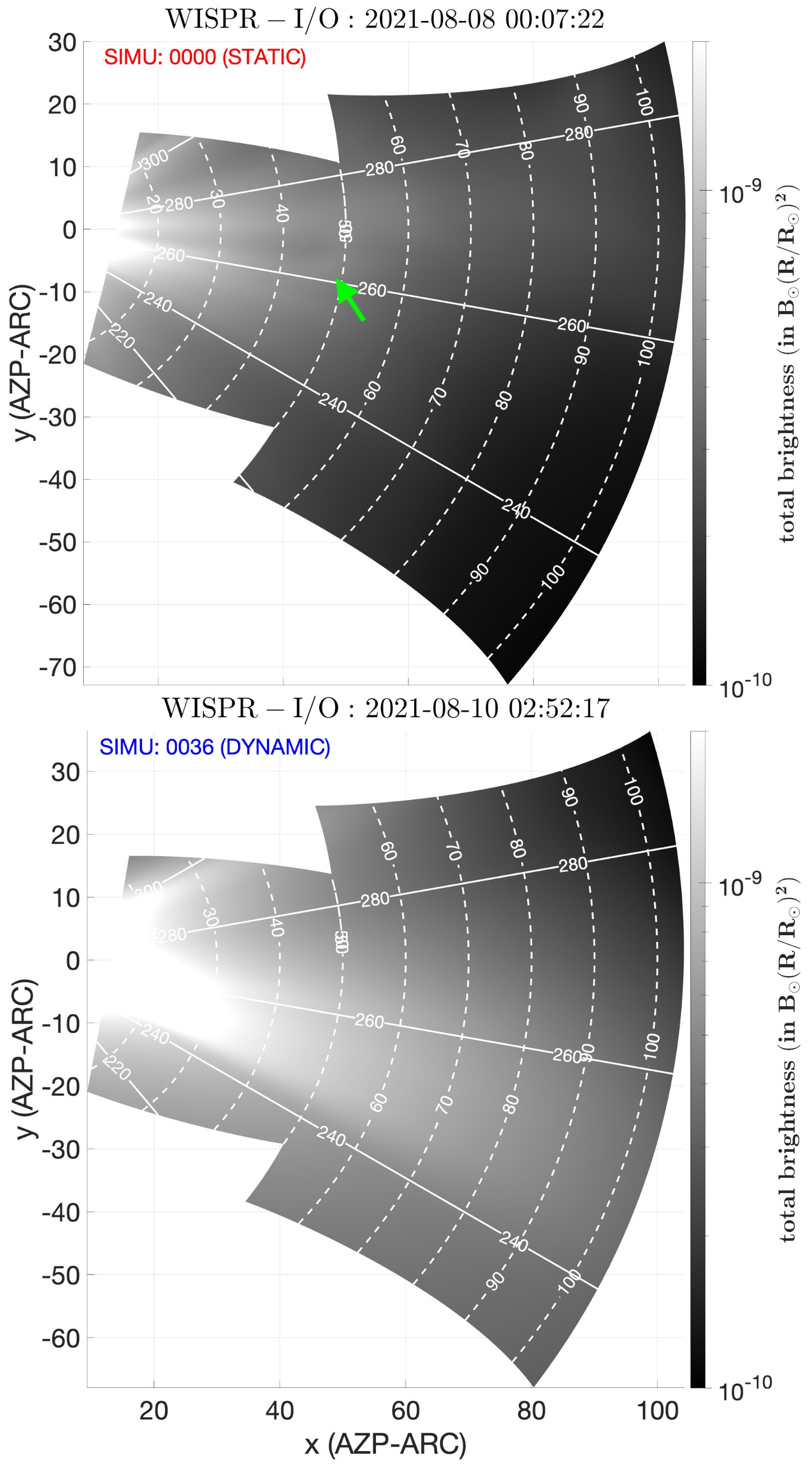}
\caption[]{ Same as Fig. \ref{fig:SYNTH_IMG_PLUTO3D}, but for   absolute brightness, plotted in units of mean solar brightness $B_\odot=2.3e7\ \rm{W.m^{-2}.sr^{-1}}$ scaled with the square of the radial distance $B_\odot(R/R_\odot)^2$, and in  logarithmic scale. An animated version of this figure is available online at \url{https://doi.org/10.5281/zenodo.8135596}.
\label{fig:SYNTH_IMG_PLUTO3D_abs}}
\end{figure*}

A similar comment can be made concerning the absolute brightness images synthesised from the fully fledged 3D modelling set-up (see Fig. \ref{fig:SYNTH_IMG_PLUTO3D_abs}), for which the results are discussed in Sect. \ref{subsec:results_3D}. Here only one flux rope structure barely stands out from the background. For a better visualisation of these transients, we decided to work primarily with difference images as those shown in the core text.

\end{appendix}

\end{document}